\documentclass{shep3}

\usepackage{axodraw,epsfig}
\usepackage{cite}
  \usepackage{graphicx}
  \usepackage{amssymb}

\preprint{
  LPSC 09/73 \\
  MZ-TH/09-19 \\
  ZU-TH 11/09
}

\title{Two-Loop Planar Corrections
to Heavy-Quark Pair Production in
the Quark-Antiquark Channel}

\author{R.~Bonciani$\rm \, ^{a, \,}$\footnote{Email: {\tt
bonciani@lpsc.in2p3.fr}},  A.~Ferroglia$\rm \, ^{b,
\,}$\footnote{Email: {\tt
ferroglia@thep.physik.uni-mainz.de}},  T.~Gehrmann$\rm \, ^{c,
\,}$\footnote{Email: {\tt Thomas.Gehrmann@physik.uzh.ch}}, and
C.~Studerus$\rm \, ^{c,\,}$\footnote{Email: {\tt cedric@physik.uzh.ch}} \\

\noindent {\it $\rm ^a$ Laboratoire de Physique Subatomique et de Cosmologie,
Universit\'e Joseph Fourier/CNRS-IN2P3/INPG, 53, rue des Martyrs,
F-38026 Grenoble, France}\\

\noindent {\it $\rm ^b$ Institut f\"ur Physik (THEP), 
Johannes Gutenberg-Universit\"at, Staudingerweg 7,
D-55099 Mainz, Germany }\\

\noindent {\it $\rm ^c$ Institut f{\"u}r Theoretische Physik,
Universit{\"a}t Z\"urich, Winterthurerstrasse 190,
CH-8057 Zurich, Switzerland}

}

\abstract{We evaluate the planar two-loop QCD diagrams contributing to the
leading color coefficient of the heavy-quark pair production cross section,  in
the quark-antiquark annihilation channel. We obtain the leading color 
coefficient in an analytic form, in terms of one- and two-dimensional harmonic 
polylogarithms of maximal weight 4. The result is valid for arbitrary  values
of the Mandelstam invariants $s$ and $t$, and of the heavy-quark mass  $m$. Our
findings agree with previous analytic results in the small-mass limit  and
numerical results for the exact amplitude. }


\begin{document}

\newcommand{\be}{\begin{equation}}
\newcommand{\ee}{\end{equation}}
\newcommand{\bfm}[1]{\mbox{\boldmath$#1$}}
\newcommand{\bff}[1]{\mbox{\scriptsize\boldmath${#1}$}}
\newcommand{\al}{\alpha}
\newcommand{\bt}{\beta}
\newcommand{\lm}{\lambda}
\newcommand{\bea}{\begin{eqnarray}}
\newcommand{\eea}{\end{eqnarray}}
\newcommand{\gm}{\gamma}
\newcommand{\Gm}{\Gamma}
\newcommand{\dl}{\delta}
\newcommand{\Dl}{\Delta}
\newcommand{\ep}{\varepsilon}
\newcommand{\vep}{\varepsilon}
\newcommand{\kp}{\kappa}
\newcommand{\Lm}{\Lambda}
\newcommand{\om}{\omega}
\newcommand{\pa}{\partial}
\newcommand{\nn}{\nonumber}
\newcommand{\dd}{\mbox{d}}
\newcommand{\grtsim}{\mbox{\raisebox{-3pt}{$\stackrel{>}{\sim}$}}}
\newcommand{\lessim}{\mbox{\raisebox{-3pt}{$\stackrel{<}{\sim}$}}}
\newcommand{\uk}{\underline{k}}
\newcommand{\gsim}{\;\rlap{\lower 3.5 pt \hbox{$\mathchar \sim$}} \raise 1pt \hbox {$>$}\;}
\newcommand{\lsim}{\;\rlap{\lower 3.5 pt \hbox{$\mathchar \sim$}} \raise 1pt \hbox {$<$}\;}
\newcommand{\Li}{\mbox{Li}}
\newcommand{\bc}{\begin{center}}
\newcommand{\ec}{\end{center}}

\def\lapprox{\lower .7ex\hbox{$\;\stackrel{\textstyle <}{\sim}\;$}}
\def\gapprox{\lower .7ex\hbox{$\;\stackrel{\textstyle >}{\sim}\;$}}

%
\newcommand{\hypF}{{}_2\mbox{F}_1}


\section{Introduction}

The top quark, with a mass of approximately $173$ GeV, is the heaviest
elementary particle  produced at colliders until now. So far, the study of the
properties of  the top-quark  was only possible at the Tevatron, where the mass
of this particle was measured with an accuracy of less than one percent. The
production cross sections and decay widths are only known with larger
uncertainties.  Because of its large mass, the top quark is expected to couple
strongly with the electroweak symmetry breaking sector. Therefore, the study of
scattering  and decay processes involving top quarks  is expected to provide
fundamental clues on the mechanism responsible for the origin of particle
masses. This is one of the main goals of the Large Hadron Collider (LHC)
physics program. 

While the Tevatron produced a few thousand top quarks, the LHC will be an
authentic top-quark factory, since it will produce millions of top quarks
already in the first low-luminosity phase \cite{Bernreuther:2008ju}. The great
wealth of data that will be obtained at the LHC will allow precise measurements
of the top-quark related observables. In turn, the latter must be matched by
equally precise calculations of the relevant cross sections and differential
distributions in perturbative QCD.

At Tevatron, top quarks are primarily produced in pairs with their
antiparticles. The same situation will be encountered at the LHC, where
experimental collaborations anticipate  measurements of the total top-quark
pair-production cross section with a  relative error between $5 \%$ and $10
\%$. On the theory side, the top-quark pair-production cross section was
calculated at next-to-leading-order (NLO) in perturbative QCD
\cite{Nason:1987xz, Nason:1989zy, Beenakker:1988bq, Beenakker:1990maa,
Mangano:1991jk, Korner:2002hy, Bernreuther:2004jv,Czakon:2008ii}. The NLO
electroweak corrections were obtained in \cite{Beenakker:1993yr, Kuhn:2005it,
Bernreuther:2006vg}. The resummation of logarithmic terms, which become large
near the production threshold, was extensively studied in 
\cite{Kidonakis:1997gm, Bonciani:1998vc, Cacciari:2003fi, Moch:2008qy,
Cacciari:2008zb, Kidonakis:2008mu, Czakon:2008cx}.

The current resummation-improved NLO predictions for the top-quark 
pair-production cross section at the LHC shows an uncertainty of about $15 \%$
\cite{Bernreuther:2008ju}; the latter is dominated by the scale uncertainty.
Therefore, in order to reduce the theoretical uncertainty at the same level of
the  expected experimental error, the calculation of the 
next-to-next-to-leading order (NNLO) perturbative QCD corrections is required.

The Feynman diagrams needed for the evaluation of the NNLO QCD corrections to
the top-quark pair production can be grouped in three categories: i) two-loop
corrections to the three-level production channels $q \bar{q} \to t\bar{t}$ and
$g g \to t \bar{t}$, ii) one-loop matrix elements with an additional parton in
the final state, and iii) tree-level matrix element with two additional partons
in the final state. The last two sets of diagrams were already evaluated in the
context of the calculation of the NLO corrections to the production of $t
\bar{t}\, +1j$~\cite{Dittmaier:2007wz}. Contributions arising from the
interference of one-loop diagrams in both the quark-antiquark and the
gluon-fusion channels were studied in \cite{Korner:2005rg, Korner:2008bn,
Anastasiou:2008vd, Kniehl:2008fd}. The first steps toward the calculation of
the  two-loop corrections  were taken in \cite{Czakon:2007ej, Czakon:2007wk};
in these papers all the relevant two-loop diagrams were evaluated in the limit
$s, |t|, |u| \gg m^2$, where $s$ is the squared center of mass energy, $t$ is
the squared momentum transfer, $u = 2 m^2-s-t$ and $m$ is the
heavy-quark mass.   A full numerical calculation of the two-loop virtual
corrections in the  $q \bar{q} \to t \bar{t}$ channel was carried out in
\cite{Czakon:2008zk}. Finally, the analytic calculation of the diagrams 
involving a closed quark loop in the  quark-antiquark channel was presented in
\cite{quarkloops}.

In this paper we describe the calculation of a conspicuous subset of the 
two-loop planar diagrams in the $q \bar{q} \to t \bar{t}$ production channel.
The calculation of these corrections allows us to obtain an analytic
expression for the leading color coefficient in the interference of the
two-loop matrix element with the tree-level amplitude. Our results are valid
for generic values of the Mandelstam invariants $s,t$ and of the heavy-quark
mass $m$. The calculation was carried out by means of a technique based on the
identification of a set of Master Integrals (MIs) through the Laporta algorithm
\cite{Laportaalgorithm}, and on their subsequent evaluation by means of the
differential equation method \cite{DiffEq}. The results are written in terms of
a suitable base of  one- and two-dimensional Harmonic Polylogarithms (HPLs)
\cite{HPLs,Vollinga,2dHPLs}.

The paper is organized as follows. In Section~\ref{nota} we introduce our
notation and conventions. In Section~\ref{sec:calc} the method employed in the
calculation is briefly described and the technical difficulties we met are
discussed. In Section~\ref{sec:renorm} we collect the explicit expression of
the counterterms needed for the ultraviolet renormalization of the Feynman 
diagrams which we evaluated.  The results which we obtained  and their
expansion near the production threshold are discussed in
Section~\ref{sec:results}. Section~\ref{sec:conc} contains our conclusions. In
the appendices we describe in some detail the HPLs we employed in the
calculation and their expansion at production threshold. Furthermore, we
collect the explicit expression of several previously unknown MIs belonging to
box topologies.


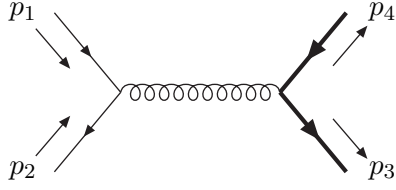
\begin{figure}
\vspace*{.5cm}
\[ \hspace*{-3mm} \vcenter{ \hbox{
  \begin{picture}(0,0)(0,0)
\SetScale{1}
  \SetWidth{.5}
\ArrowLine(-55,30)(-30,0)
\ArrowLine(-30,0)(-55,-30)
\Gluon(-30,0)(30,0){3}{10}
\LongArrow(-62,24)(-50,10)
\LongArrow(-62,-24)(-50,-10)
\LongArrow(50,10)(62,24)
\LongArrow(50,-10)(62,-24)
  \SetWidth{1.6}
\ArrowLine(55,30)(30,0)
\ArrowLine(30,0)(55,-30)

\Text(-67,27)[cb]{$p_1$}
\Text(-67,-33)[cb]{$p_2$}
\Text(69,-33)[cb]{$p_3$}
\Text(69,27)[cb]{$p_4$}
\end{picture}}
}
\]
\vspace*{.6cm} 
\caption{\it Tree-level  amplitude. Massive
quarks are indicated by a thick line.} 
\label{figTree}
\end{figure}


\section{Notation and Conventions \label{nota}}

We consider the scattering process \be q(p_1) + \overline{q}(p_2)
\longrightarrow  t(p_3) + \overline{t}(p_4) \, , \ee
in Euclidean kinematics, where $p_i^2 = 0$ for $i=1,2$  and
$p_j^2 = -m^2$ for  $j=3,4$. The Mandelstam variables are defined as
follows
\be
s = -\left(p_1 + p_2 \right)^2 \, , \quad
t = -\left(p_1 - p_3 \right)^2 \, , \quad
u = -\left(p_1 - p_4 \right)^2 \, .
\ee
Conservation of momentum implies that $s +t +u = 2 m^2$.

The squared matrix element (averaged over the spin and color of the
incoming quarks and summed over the spin of the outgoing ones), calculated in
$d = 4 -2 \varepsilon$ dimensions, can be expanded in powers of the strong
coupling constant $\alpha_S$ as follows:
\be \label{M2}
|\mathcal{M}|^2(s,t,m,\varepsilon) = \frac{4 \pi^2 \alpha_S^2}{N_c^2}
\left[{\mathcal A}_0 +
\left(\frac{\alpha_S}{ \pi} \right) {\mathcal A}_1 +
\left(\frac{\alpha_S}{ \pi} \right)^2 {\mathcal A}_2 +
{\mathcal O}\left( \alpha_S^3\right)\right] \, .
\ee
The tree-level amplitude involves a single diagram (Fig.~\ref{figTree}) and its contribution to Eq.~(\ref{M2}) is given by
\be
{\mathcal A}_0=  4  N_c \, C_F
\left[ \frac{(t-m^2)^2+(u-m^2)^2}{s^2} + \frac{2 m^2}{s} - \varepsilon\right] 
\, ,
\label{treeM}
\ee
where  $N_c$ is the number of colors and  $C_F = (N_c^2-1)/2N_c$.

The ${\mathcal O}(\alpha_S)$ term ${\mathcal A}_1$  in Eq.~(\ref{M2}) arises 
from the interference of one-loop diagrams with the tree-level amplitude
\cite{Nason:1987xz,Nason:1989zy,Beenakker:1988bq,Beenakker:1990maa,Mangano:1991jk,Korner:2002hy,Bernreuther:2004jv}.
The ${\mathcal O}(\alpha_S^2)$ term ${\mathcal A}_2$ consists of two parts, the
interference of two-loop diagrams with the Born amplitude and the interference 
of one-loop diagrams among themselves:
\begin{displaymath}
{\mathcal A}_2 = {\mathcal A}_2^{(2\times 0)} + {\mathcal A}_2^{(1\times 1)}\;.
\end{displaymath}
The latter term ${\mathcal A}_2^{(1\times 1)}$ was
studied  extensively  in
\cite{Korner:2005rg,Korner:2008bn}. ${\mathcal A}_2^{(2\times 0)}$, originating
from the two-loop  diagrams, can be decomposed according to color and
flavor structures as follows:
\bea  {\mathcal A}_2^{(2\times 0)}  &=&  N_c C_F \Biggl[ N_c^2 A
+B +\frac{C}{N_c^2}  + N_l \left( N_c D_l + \frac{E_l}{N_c}
\right)
+ N_h \left( N_c D_h + \frac{E_h}{N_c} \right)  \nn \\
& & \hspace*{10mm} + N_l^2 F_l + N_l N_h F_{lh} + N_h^2 F_h\Biggr] \, ,
\label{colstruc}
\eea
where $N_l$ and $N_h$ are the number of light- and heavy-quark flavors,
respectively. The coefficients $A,B,\ldots,F_h$ in Eq.~(\ref{colstruc}) are
functions of $s$, $t$, and $m$, as well as of the dimensional regulator
$\varepsilon$. These quantities were calculated in  \cite{Czakon:2007ej} in the
approximation $s,|t|,|u| \gg m^2$. For a fully differential description of top
quark pair production at NNLO, the complete mass dependence of  ${\mathcal
A}_2^{(2\times 0)}$ is required. An exact numerical expression for it has been
obtained  in \cite{Czakon:2008zk}.  Exact analytic formulae for all the
coefficients arising from Feynman diagrams involving closed quark loops
($D_l,E_l,D_h,E_h,F_l,F_h,F_{lh}$)  were obtained in \cite{quarkloops}.  In
this work, we provide an exact analytic  expression for the coefficient  $A$
in  Eq.~(\ref{colstruc}), which arises from planar Feynman diagrams only.

\section{Calculation\label{sec:calc}}


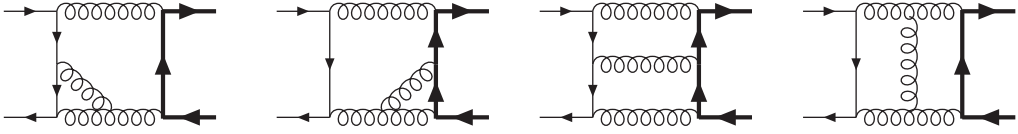
\begin{figure}
\vspace*{.5cm}
\[ \hspace*{-3mm}
\vcenter{
\hbox{\begin{picture}(0,0)(0,0)
\SetScale{1}
  \SetWidth{.5}
  \ArrowLine(-40,20)(-20,20)
  \ArrowLine(-20,0)(-20,-20)
  \ArrowLine(-20,20)(-20,0)
  \ArrowLine(-20,-20)(-40,-20)
  \Gluon(-20,20)(20,20){3}{7.5}
  \Gluon(-20,-20)(20,-20){3}{7.5}
  \Gluon(-20,0)(0,-18){3}{4.5}
  \SetWidth{1.6}
  \ArrowLine(20,20)(40,20)
  \ArrowLine(20,-20.5)(20,20.5)
  \ArrowLine(40,-20)(20,-20)
\end{picture} }
}
\hspace{3.5cm}
  \vcenter{
\hbox{\begin{picture}(0,0)(0,0)\SetScale{1}
  \SetWidth{.5}
  \ArrowLine(-40,20)(-20,20)
  \ArrowLine(-20,20)(-20,-20)
  \ArrowLine(-20,-20)(-40,-20)
  \Gluon(-20,20)(20,20){3}{7.5}
  \Gluon(-20,-20)(20,-20){3}{7.5}
  \Gluon(0,-18)(20,0){3}{4.5}
  \SetWidth{1.6}
  \ArrowLine(20,20)(40,20)
  \ArrowLine(20,-20.5)(20,0)
  \ArrowLine(20,0)(20,20.5)
  \ArrowLine(40,-20)(20,-20)
\end{picture}}
}
\hspace{3.5cm}
  \vcenter{
\hbox{\begin{picture}(0,0)(0,0)
\SetScale{1}
  \SetWidth{.5}
  \ArrowLine(-40,20)(-20,20)
  \ArrowLine(-20,20)(-20,0)
  \ArrowLine(-20,0)(-20,-20)
  \ArrowLine(-20,-20)(-40,-20)
  \Gluon(-20,20)(20,20){3}{7.5}
  \Gluon(-20,-20)(20,-20){3}{7.5}
  \Gluon(-20,0)(20,0){3}{7.5}
  \SetWidth{1.6}
  \ArrowLine(20,20)(40,20)
  \ArrowLine(20,-20.5)(20,0)
  \ArrowLine(20,0)(20,20.5)
  \ArrowLine(40,-20)(20,-20)
\end{picture}}
}
\hspace{3.5cm}
  \vcenter{
\hbox{\begin{picture}(0,0)(0,0)
\SetScale{1}
  \SetWidth{.5}
  \ArrowLine(-40,20)(-20,20)
  \ArrowLine(-20,20)(-20,-20)
  \ArrowLine(-20,-20)(-40,-20)
  \Gluon(-20,20)(20,20){3}{7.5}
  \Gluon(-20,-20)(20,-20){3}{7.5}
  \Gluon(0,18)(0,-18){3}{5.5}
  \SetWidth{1.6}
  \ArrowLine(20,20)(40,20)
  \ArrowLine(20,-20.5)(20,20.5)
  \ArrowLine(40,-20)(20,-20)
\end{picture}}
}
\]
\vspace*{.3cm}
\caption{\it Some of the two-loop planar box diagrams involved in the 
calculation.}
\label{boxes}
\end{figure}


The package QGRAF~\cite{qgraf} generates 218 two-loop Feynman diagrams
contributing to the process $q\bar q\to t\bar t$.  After treating the QGRAF
output with FORM~\cite{FORM} in order to carry out the color (and Dirac)
algebra,  one finds that there are 44 non-vanishing  diagrams
contributing to the leading color coefficient in Eq.~(\ref{colstruc}). Some of
the box diagrams involved in the calculation are shown in Fig.~\ref{boxes}. All
the two-loop graphs encountered in our calculation can be treated by 
employing the technique based upon the Laporta algorithm for the
reduction to a set of Master Integrals (MIs). The MIs are then evaluated by 
means of the differential equation method.

A subset of the  MIs needed in the calculation were available in the 
literature~\cite{Argeri:2002wz,Bonciani:2003te,Aglietti:2003yc,DK,Aglietti:2004tq,CGR}.
They were employed in previous two-loop calculations of the heavy-quark form
factors~\cite{Bernreuther:2004ih}, amplitudes for Bhabha
scattering~\cite{electronloop},  heavy-to-light quark transitions
\cite{heavytolight}, and in the calculation  of the fermionic corrections to
$q \bar{q} \to t \bar{t}$~\cite{quarkloops}.  The MIs not included in this 
subset are a part of the original findings of this paper.

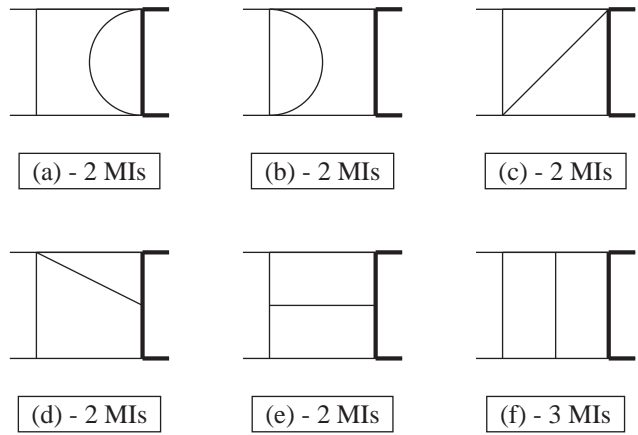
\begin{figure}
\vspace*{.5cm}
\[ \hspace*{-3mm}
  \vcenter{
\hbox{\begin{picture}(0,0)(0,0)
\SetScale{1}
  \SetWidth{.5}
  \Line(-30,20)(30,20)
  \Line(-30,-20)(30,-20)
  \Line(-20,20)(-20,-20)
  \Line(-20,20)(20,20)
\CArc(20,0)(20,90,270)
\BText(0,-41){(a) - 2 MIs}
  \SetWidth{1.6}
  \Line(20,20)(30,20)
  \Line(20,-20)(30,-20)
  \Line(20,20)(20,-20)
%
\end{picture}}
}
\hspace{3.1cm}
\vcenter{
\hbox{\begin{picture}(0,0)(0,0)

\SetScale{1}
  \SetWidth{.5}
  \Line(-30,20)(30,20)
  \Line(-30,-20)(30,-20)
  \Line(-20,20)(-20,-20)
  \Line(-20,20)(20,20)
\CArc(-20,0)(20,270,450)
\BText(0,-41){(b) - 2 MIs}
  \SetWidth{1.6}
  \Line(20,20)(30,20)
  \Line(20,-20)(30,-20)
  \Line(20,20)(20,-20)
\end{picture}}
}
\hspace{3.1cm}
 \vcenter{\hbox{\begin{picture}(0,0)(0,0)
\SetScale{1}
  \SetWidth{.5}
  \Line(-30,20)(30,20)
  \Line(-30,-20)(30,-20)
  \Line(-20,20)(-20,-20)
  \Line(-20,20)(20,20)
  \Line(-20,-20)(20,20)
\BText(0,-41){(c) - 2 MIs}
  \SetWidth{1.6}
  \Line(20,20)(30,20)
  \Line(20,-20)(30,-20)
  \Line(20,20)(20,-20)
\end{picture}}} 
\]
\vspace*{1.8cm}
\[ \hspace*{-3mm}
\vcenter{\hbox{\begin{picture}(0,0)(0,0)
\SetScale{1}
  \SetWidth{.5}
  \Line(-30,20)(30,20)
  \Line(-30,-20)(30,-20)
  \Line(-20,20)(-20,-20)
  \Line(-20,20)(20,20)
  \Line(-20,20)(20,0)
\BText(0,-41){(d) - 2 MIs}
  \SetWidth{1.6}
  \Line(20,20)(30,20)
  \Line(20,-20)(30,-20)
  \Line(20,20)(20,-20)
\end{picture}}
  } 
\hspace{3.1cm}
 \vcenter{\hbox{\begin{picture}(0,0)(0,0)
\SetScale{1}
  \SetWidth{.5}
  \Line(-30,20)(30,20)
  \Line(-30,-20)(30,-20)
  \Line(-20,20)(-20,-20)
  \Line(-20,20)(20,20)
  \Line(-20,0)(20,0)
\BText(0,-41){(e) - 2 MIs}
  \SetWidth{1.6}
  \Line(20,20)(30,20)
  \Line(20,-20)(30,-20)
  \Line(20,20)(20,-20)
\end{picture}}
  }
\hspace{3.1cm}
 \vcenter{\hbox{\begin{picture}(0,0)(0,0)

\SetScale{1}
  \SetWidth{.5}
  \Line(-30,20)(30,20)
  \Line(-30,-20)(30,-20)
  \Line(-20,20)(-20,-20)
  \Line(-20,20)(20,20)
  \Line(0,-20)(0,20)
\BText(0,-41){(f) - 3 MIs}
  \SetWidth{1.6}
  \Line(20,20)(30,20)
  \Line(20,-20)(30,-20)
  \Line(20,20)(20,-20)
\end{picture}}
  }
\]
\vspace*{.8cm}
\caption{\it New non-reducible topologies encountered in the calculation
of the planar diagrams.  The number of Master Integrals related to each topology is indicated in the figure. 
}
\label{nlMIs}
\end{figure}

The reduction to MIs was carried out with an implementation of  the Laporta
algorithm written in {\tt C++}~\cite{Cedriccode}, and large parts of it were
cross checked with the {\tt Maple} package {\tt A.I.R.}
\cite{Anastasiou:2004vj}.  The six so far unknown irreducible topologies
encountered in the calculation of the planar diagrams  are shown in
Fig.~\ref{nlMIs}.   In the figure, thick internal  lines indicate massive
propagators, while thin lines indicate massless ones. An external dashed  leg
carries a squared momentum $ (p_1+p_2)^2 = -s$; other external  lines indicate
particles on their mass-shell, where $p_i^2 = 0$ for thin lines and $p_i^2 =
-m^2$ for thick lines.
The MIs in Fig.~\ref{nlMIs} are collected in Appendix~\ref{AppMIs} and in a 
file included with the arXiv submission of the present paper.

In calculating the MIs by means of the differential equation method, it is
crucial to fix the undetermined integration constant(s) through a boundary 
condition. While there is no general method available to fix this boundary
condition, it is usually sufficient to know the behavior of the MI at some
particular kinematic point. For example, knowing that the integral is regular
for a certain value of $s$, one can impose the regularity of the solution of the
differential equation at that point.  In cases in which considerations related
to the regularity of the  solutions of the differential equations are not
sufficient to fix all of the  integration constants, it is necessary to directly
evaluate the MIs at a specific phase-space point. This can be done using
techniques based on Mellin-Barnes representations of the integrals~\cite{MB},
with the help of  the {\tt Mathematica} packages {\tt Ambre} \cite{Gluza:2007rt}
and {\tt MB} \cite{Czakon:2005rk}. In order to numerically check  the analytic
calculation of the MIs, we employed the sector decomposition
technique~\cite{secdec}, implemented in the {\tt Mathematica} package {\tt
FIESTA}~\cite{FIESTA}.

All the MIs were calculated in the non-physical region $s<0$, where they are
real and can be conveniently written as functions of the dimensionless
variables
\be
x = \frac{\sqrt{-s+4 m^2} - \sqrt{-s}}{\sqrt{-s+4 m^2} + \sqrt{-s}} \, ,
\qquad y = \frac{-t}{m^2}  \, .
\ee
The transcendental functions appearing in the results are one- and
two-dimensional harmonic polylogarithms
(HPLs)~\cite{HPLs,Vollinga,2dHPLs,Aglietti:2004tq} of maximum weight four. In
Appendix~\ref{AppGPLs}, we briefly review the definition of  the HPLs appearing
in the calculation. All the HPLs appearing in the  analytic expression of the
coefficient $A$ can be evaluated numerically with arbitrary precision by
employing the methods and codes described  in~\cite{Vollinga}.

Following the procedure outlined in the present section, it is possible to
obtain the expression of the bare squared matrix elements involving planar
two-loop diagrams. The renormalization of the ultraviolet divergencies is
discussed in the next section.


\section{Renormalization\label{sec:renorm}}

The renormalized QCD matrix element is obtained from the bare one by expanding
the following expression :
\be
{\mathcal A}_{{\mbox{\small ren}}} = \prod_{n} Z^{1/2}_{{\mbox{\tiny WF}},n} \,
{\mathcal A}_{{\mbox{\small bare}}}
\left( \alpha_{S, {\mbox{\small bare}}} \to Z_{\alpha_S} \alpha_S \, ,
m_{{\mbox{\small bare}}} \to Z_m m \right) \, ,
\label{renM}
\ee
where $Z_{{\mbox{\tiny WF}},n}$ is the external leg wave function
renormalization factor, $\alpha_S$ is the renormalized coupling constant and
$m$ is the renormalized heavy-quark mass. (In the rest of the section we
suppress the subscript ``$S$'' in $\alpha_S$).

We postpone the discussion of mass renormalization to the end of the section
and we start by  considering the coupling constant and wave function
renormalization.

We introduce the following quantities:
\be
a_0 = \frac{\alpha_{\mbox{\small bare}}}{\pi}  \, ,
\qquad \mbox{and} \qquad
a = \frac{\alpha}{\pi}  \, .
\ee
By expanding the amplitude and the wave function renormalization factor
in $a_0$ we find:
\bea
{\mathcal A}_{{\mbox{\small ren}}}(\alpha_{\mbox{\small bare}}) &=&
    a_0 {\mathcal A}_0
  + a_0^2 {\mathcal A}_1
  + a_0^3 {\mathcal A}_2
  + {\mathcal O}(a_0^4) \, , \nn \\
Z_{{\mbox{\tiny WF}},n} & = & 1
  + a_0 \delta Z^{(1)}_{{\mbox{\tiny WF}},n}
  + a_0^2 \delta Z^{(2)}_{{\mbox{\tiny WF}},n}
  + {\mathcal O}(a_0^3) \, .\label{exp1}
\eea
The relation between $a_0$ and $a$ is given by:
\be
a_0  =  a
  + a^2 \delta Z^{(1)}_{\alpha}
  + a^3\delta Z^{(2)}_{\alpha}
  + {\mathcal O}(a^4) \, . \label{exp2}
\ee
By employing  Eqs.~(\ref{exp1},\ref{exp2}) in Eq.~(\ref{renM}) we find
\bea
{\mathcal A}_{{\mbox{\small ren}}} & = & a {\mathcal A}_0
+ a^2 {\mathcal A}^{(1)} _{{\mbox{\small ren}}} +
a^3 {\mathcal A}^{(2)} _{{\mbox{\small ren}}} + \mathcal{O}(a^4) \, , \nn \\
{\mathcal A}^{(1)} _{{\mbox{\small ren}}} & = & {\mathcal A}_1 +
\left(\sum_n \frac{1}{2} \delta Z^{(1)}_{{\mbox{\tiny WF}},n} + \delta
Z^{(1)}_{\alpha}\right)  {\mathcal A}_0 \, , \nn \\
{\mathcal A}^{(2)} _{{\mbox{\small ren}}} & = &  {\mathcal A}_2 +
 \left(\sum_n \frac{1}{2} \delta Z^{(1)}_{{\mbox{\tiny WF}},n}  + 2 \delta
Z^{(1)}_{\alpha} \right) {\mathcal A}_1+
\left(-\sum_n \frac{1}{8} \left(\delta Z^{(1)}_{{\mbox{\tiny WF}},n}\right)^2  
\right.   \nn \\
& & 
+ \left.\sum_n \frac{1}{2} \delta Z^{(2)}_{{\mbox{\tiny WF}},n} 
+ \delta Z^{(1)}_{\alpha}\sum_n \delta Z^{(1)}_{{\mbox{\tiny WF}},n}
+ \delta Z^{(2)}_{\alpha}
\right) {\mathcal A}_0 \, .
\label{exprenM}
\eea
In the equations above, ${\mathcal A}_i$ represents the bare amplitude at $i$
loops stripped of the factor $a$. In the case of the process $ q \overline{q}
\to t \overline{t}$, the wave function renormalization factors of  massless
quarks vanish at one loop, while the ones of the massive quarks in the on-shell
renormalization scheme are given by
\be
\delta Z^{(1)}_{{\mbox{\tiny WF}},M} = C(\vep) \,
             \left(\frac{\mu^2}{m^2} \right)^{\vep} C_F
         \left( -\frac{3}{4 \ep} - \frac{1}{1-2 \ep}  \right) \, ,
\ee
where the subscript $M$ indicates massive quarks and where $C(\vep) = (4 
\pi)^{\vep} \Gamma(1+\vep)$.
The one-loop renormalization constant for $\alpha$  in the $\overline{MS}$
scheme is given by
\be
\delta Z^{(1)}_{\alpha} = C(\vep) \,
\frac{e^{-\gamma \vep}}{\Gamma(1+\vep)} \left( -\frac{\beta_0}{2 \vep}\right) \, ,
\ee
where $\beta_0 = 11/6 C_A  -1/3 (N_l + N_h)$ and $\gamma$ is the
Euler-Mascheroni constant $\gamma \approx  0.577216$.

Therefore, the overall one-loop counter term is given by
\be
\delta Z^{(1)}_{{\mbox{\tiny WF}},M} + \delta Z^{(1)}_{\alpha} =
-\frac{C(\vep)}{4 \ep} \left[ 
2 \beta_0 + \left( 3 + 4 \vep + 3 \vep \ln{\left(
\frac{\mu^2}{m^2}\right)} \right) C_F  
+ {\mathcal O} \left( \vep^2 \right) 
\right] \, .
\ee
%

\begin{figure}
\vspace*{.5cm}
\[ \hspace*{-3mm}
  \vcenter{
\hbox{\begin{picture}(0,0)(0,0)
\SetScale{.8}
  \SetWidth{.5}
\ArrowLine(-55,30)(-30,0) \ArrowLine(-30,0)(-55,-30)
\Gluon(-30,0)(-15,0){3}{2.5} \Gluon(15,0)(30,0){3}{2.5}
\GlueArc(0,0)(15,0,180){3}{8} \GlueArc(0,0)(15,180,360){3}{8}
\BText(0,-45){1l-1}
  \SetWidth{1.6}
\ArrowLine(30,0)(55,30)
\ArrowLine(55,-30)(30,0)
\end{picture}}
}
%
\hspace*{3.7cm}
%
\vcenter{
\hbox{ \begin{picture}(0,0)(0,0)
\SetScale{.8}
  \SetWidth{.5}
\ArrowLine(-55,30)(-30,0) \ArrowLine(-30,0)(-55,-30)
\Gluon(-30,0)(-15,0){3}{2.5} \Gluon(15,0)(30,0){3}{2.5}
\DashArrowArcn(0,0)(15,0,180){3}
\DashArrowArcn(0,0)(15,180,360){3}
\BText(0,-45){1l-2}
  \SetWidth{1.6}
\ArrowLine(30,0)(55,30)
\ArrowLine(55,-30)(30,0)
\end{picture}}
}
\hspace{3.7cm}
  \vcenter{
\hbox{\begin{picture}(0,0)(0,0)
\SetScale{.8}
  \SetWidth{.5}
\Gluon(30,0)(0,0){3}{4.5} \ArrowLine(-40,30)(-40,-30)
\ArrowLine(-55,30)(-40,30) \ArrowLine(-40,-30)(-55,-30)
\Gluon(-40,30)(0,0){3}{7.5} \Gluon(0,0)(-40,-30){3}{7.5}
\BText(0,-45){1l-3}
  \SetWidth{1.6}
\ArrowLine(30,0)(55,30)
\ArrowLine(55,-30)(30,0)
\end{picture}}
}
\hspace{3.7cm}
  \vcenter{
\hbox{\begin{picture}(0,0)(0,0)
\SetScale{.8}
  \SetWidth{.5}
\Gluon(30,0)(0,0){3}{4.5} \Gluon(-40,-30)(-40,30){3}{10.5}
\ArrowLine(-55,30)(-40,30) \ArrowLine(-40,-30)(-55,-30)
\ArrowLine(-40,30)(0,0) \ArrowLine(0,0)(-40,-30)
\BText(0,-45){1l-4}
  \SetWidth{1.6}
\ArrowLine(30,0)(55,30)
\ArrowLine(55,-30)(30,0)
\end{picture}}
}
\]
\vspace*{1.5cm}
\[
\vcenter{
\hbox{\begin{picture}(0,0)(0,0)
\SetScale{.8}
  \SetWidth{.5}
\ArrowLine(-55,30)(-30,0) \ArrowLine(-30,0)(-55,-30)
\Gluon(-30,0)(0,0){3}{4.5} \Gluon(0,0)(40,30){3}{7.5}
\Gluon(0,0)(40,-30){3}{7.5} \BText(0,-45){1l-5}
  \SetWidth{1.6}
\ArrowLine(40,30)(55,30)
\ArrowLine(55,-30)(40,-30)
\ArrowLine(40,-30)(40,30)
\end{picture}}
}
\hspace{3.8cm}
\vcenter{
\hbox{\begin{picture}(0,0)(0,0)
\SetScale{.8}
  \SetWidth{.5}
\ArrowLine(-55,30)(-30,0) \ArrowLine(-30,0)(-55,-30)
\Gluon(-30,0)(0,0){3}{4.5} \Gluon(40,-30)(40,30){3}{10.5}
\BText(0,-45){1l-6}
  \SetWidth{1.6}
\ArrowLine(40,30)(55,30)
\ArrowLine(55,-30)(40,-30)
\ArrowLine(0,0)(40,30)
\ArrowLine(40,-30)(0,0)
\end{picture}}
  }
%
\hspace*{3.8cm}
%
\vcenter{\hbox{\begin{picture}(0,0)(0,0)
\SetScale{.8}
  \SetWidth{.5}
\ArrowLine(-40,30)(-40,-30) \ArrowLine(-55,30)(-40,30)
\ArrowLine(-40,-30)(-55,-30) \Gluon(-40,30)(40,30){3}{12.5}
\Gluon(40,-30)(-40,-30){3}{12.5} \BText(0,-50){1l-7}
  \SetWidth{1.6}
\ArrowLine(40,-30)(40,30)
\ArrowLine(40,30)(55,30)
\ArrowLine(55,-30)(40,-30)
\end{picture}}}
\hspace{3.8cm}
\vcenter{\hbox{\begin{picture}(0,0)(0,0)
\SetScale{.8}
  \SetWidth{.5}
\ArrowLine(-40,30)(-40,-30)
\ArrowLine(-55,30)(-40,30)
\ArrowLine(-40,-30)(-55,-30)

\Gluon(-40,30)(40,-30){3}{15.5} \Gluon(-40,-30)(40,30){3}{15.5}
\BText(0,-50){1l-8}
  \SetWidth{1.6}
\ArrowLine(40,-30)(40,30)
\ArrowLine(40,30)(55,30)
\ArrowLine(55,-30)(40,-30)
\end{picture}}
  }
\]
\vspace*{.8cm}
\caption{\it One-loop diagrams (excluding the diagrams with closed quark loops). 
Thin arrow lines represent massless quarks, thick arrow line massive quarks, 
dashed arrow lines are Faddeev-Popov ghosts, and coiled lines are gluons.}
\label{fig1ltot}
\end{figure}
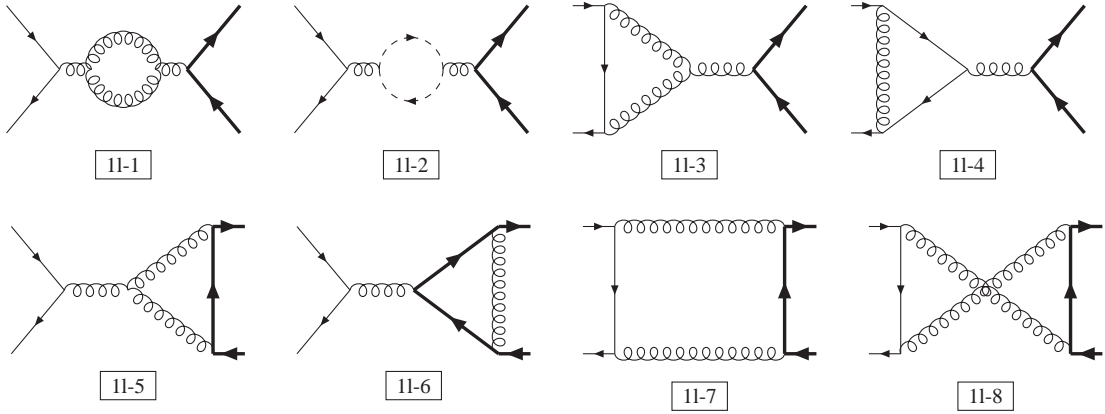

Here we are interested in finding the ultraviolet counterterm needed to
renormalize the two-loop diagrams not involving closed quark loops. The part of
this counterterm needed to renormalize the leading color structure in
Eq.~(\ref{colstruc}) can then be trivially extracted.
Taking into account the fact that the wave function renormalization factors are
zero for the incoming particles and identical for the massive ones, one finds
that the renormalized amplitude (excluding quark-loop diagrams) is
\bea
{\mathcal A}^{(2,{\tt g})} _{{\mbox{\small ren}}} &=&  {\mathcal A}_2^{({\tt g})} +
 \left(\delta Z^{(1)}_{{\mbox{\tiny WF}},M}  + 2 \delta
Z^{(1,C_A)}_{\alpha} \right)\sum_{j=1}^{8} \nn {\mathcal A}^{(d_j)}_1-
 \delta Z_m^{(1)}\sum_{l=5}^{8} {\mathcal A}^{(d_l,\mbox{{\tiny {\tt mass CT }}})}_1 \\
& & + \left(\delta Z^{(2,{\tt g})}_{{\mbox{\tiny WF}},M} +
  2 \delta Z^{(1,C_A)}_{\alpha}\delta Z^{(1)}_{{\mbox{\tiny WF}},M}
 +\delta Z^{(2,{\tt g})}_{\alpha}
\right) {\mathcal A}_0\, .
\label{renNl}
\eea
In Eq.~(\ref{renNl}), the quantity ${\mathcal A}^{(d_j)}_1$ is the amplitude of
the $j$-th diagram in Fig.~\ref{fig1ltot} (stripped of the factor $a$). The
quantity ${\mathcal A}^{(d_l,\mbox{{\tiny {\tt mass CT }}})}$ indicates the
$l$-th diagram in Fig.~\ref{fig1ltot} with a mass counter term insertion in one
of the internal heavy quark lines. The renormalization coefficients not
previously defined are:
\bea
\delta Z^{(1)}_{m} &=&  \delta Z^{(1)}_{{\mbox{\tiny WF}},M}\, , \nn \\
\delta Z^{(1,C_A)}_{\alpha} &=& -C(\vep) \, \frac{e^{-\gamma \vep}}{\Gamma(1+\vep)} \, C_A
\frac{11}{12 \vep}  \, , \nn \\
\delta Z^{(2,{\tt g})}_{\alpha} &=& C(\vep)^2 \, \left( \frac{e^{-\gamma \vep}}{\Gamma(1+\vep)}
\right)^2 \frac{1}{4 \vep} \left[
\left(\frac{11}{6}\right)^2\frac{C_A^2}{\vep} - \frac{17}{12} C_A^2
 \right] \, , \nn \\
\delta Z^{(2,{\tt g})}_{{\mbox{\tiny WF}},M} &=& C(\vep)^2 \left(\frac{\mu^2}{m^2} \right)^{2 \vep} \, C_F \Biggr[
C_F \left(\frac{9}{32\vep^2} + \frac{51}{64 \vep} + \frac{433}{128} -\frac{3}{2} \zeta(3)  
+ \pi^2 \ln{2} - \frac{13}{16} \pi^2 \right) \nn \\
&& + C_A\left(-\frac{11}{32 \vep^2}-\frac{101}{64 \vep}-
\frac{803}{128} + \frac{3}{4} \zeta(3) - \frac{\pi^2}{2} \ln{2} + \frac{5}{16}\pi^2 \right)
 +
   {\mathcal O} \left( \ep\right)\Biggl] \label{rencoeff} \, ,
\eea
and they can be found in~\cite{Czakon:2007ej,Melnikov:2000zc}.

\section{Results\label{sec:results}}

The main result of the present paper is an analytic, non-approximated 
expression for the coefficient $A$ in Eq.~(\ref{colstruc}).  Since such a
result is too long to be explicitly printed here, we included in the arXiv
submission of this work a text file with the complete result, which is written
in terms of one- and two-dimensional HPLs of maximum weight four. Since the
coefficients in Eq.~(\ref{colstruc}) still contain infrared poles, the result
is dependent on the choice of a global, $\vep$-dependent normalization factor.
With our choice, we factor out an overall coefficient
\be
C^2(\vep) = \bigl[\left(4 \pi \right)^{\vep} \Gamma(1+\vep) \bigr]^2 \, .
\label{Cep}
\ee
We also provide a code that numerically evaluates  the analytic expression of
the quantities listed above for arbitrary values of the mass scales involved in
the calculation. The code is written in {\tt C++} and uses the package for
the evaluation of multiple polylogarithms within  {\tt GiNaC}~\cite{Vollinga}.

\begin{figure}
\centering
\begin{tabular}{cc}
\includegraphics[scale=.6]{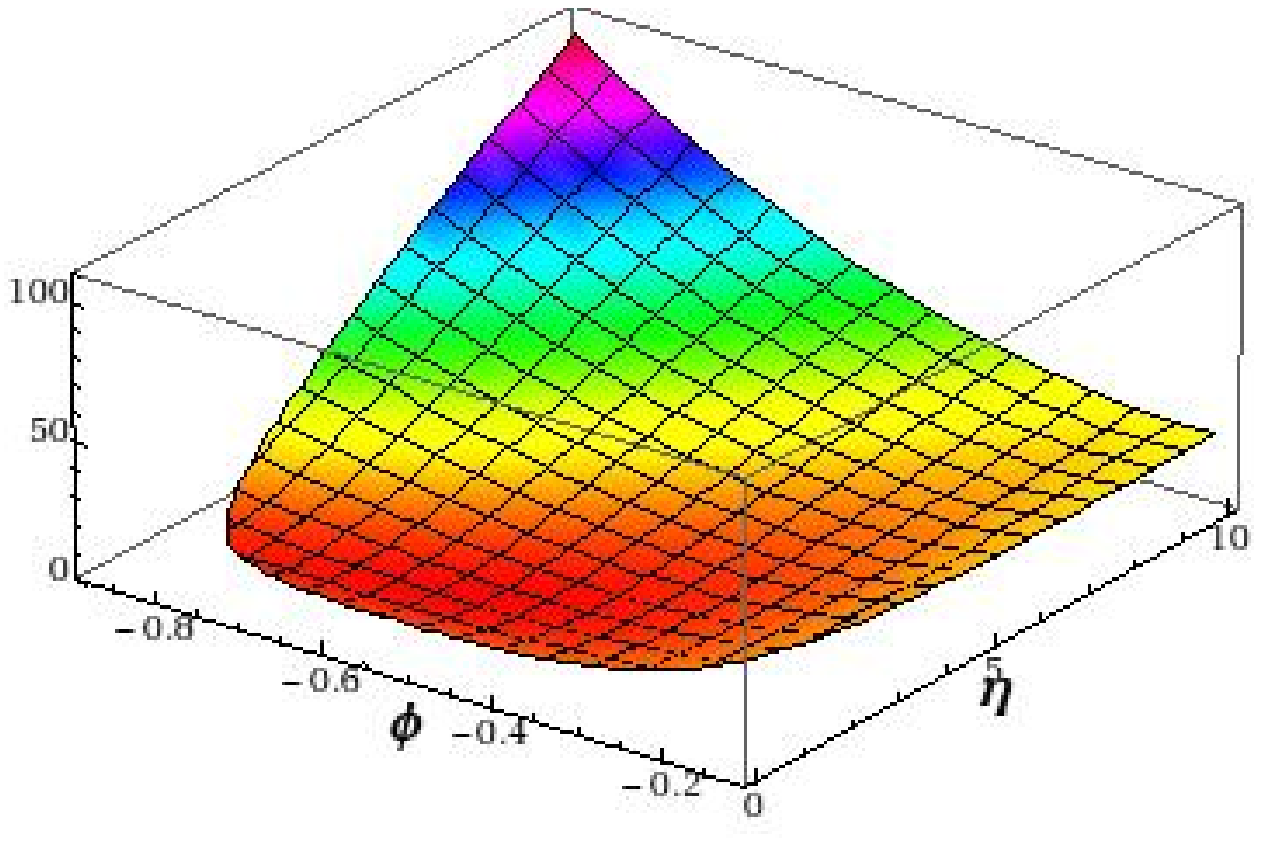} 
\includegraphics[scale=.6]{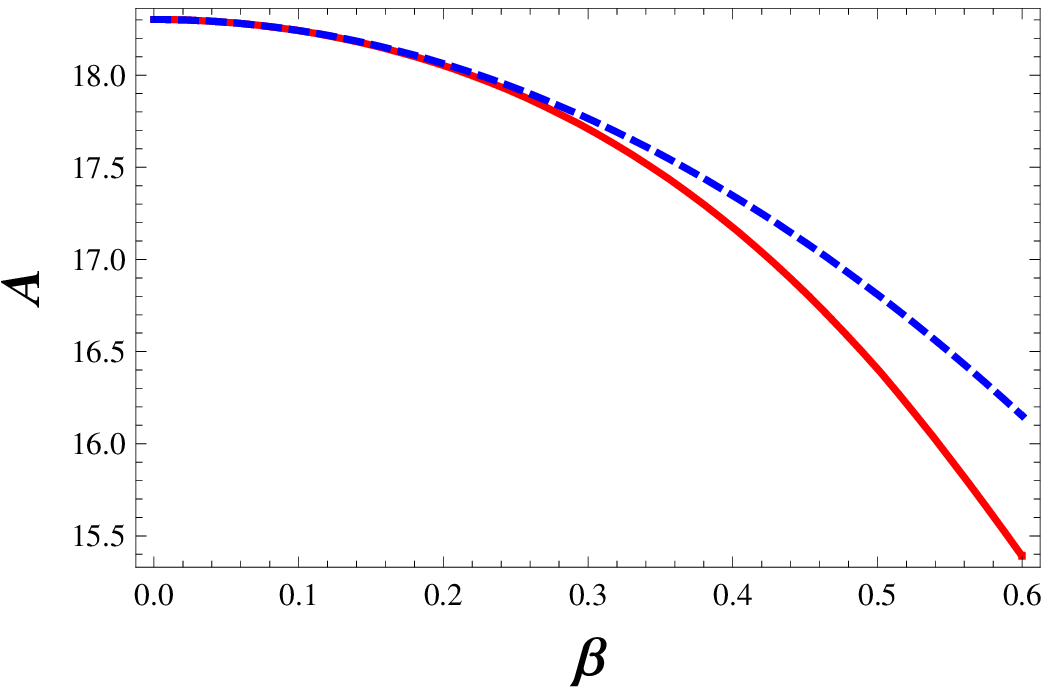}
\end{tabular}
\caption{\it Left: finite part of the coefficient $A$ as a function of the variables
$\eta$ and $\phi$. Right: the exact value of  $A^{(0)}$ as a function of
$\beta$ (red solid curve) versus its expansion close to threshold up to terms
of order $\beta^2$ (blue dashed curve).   Both curves are plotted for $\xi =
1/2$. In both cases we used the normalization adopted in \cite{Czakon:2008zk} 
to facilitate comparisons. }
\label{surface}
\end{figure}

In order to cross check our results, we expanded them in the  $s,|t|,|u| \gg m^2$
limit. The first term in the expansion agrees with the results published in
\cite{Czakon:2007ej}; the second order term agrees with the results found in
the {\tt Mathematica} files  included in the arXiv version of
\cite{Czakon:2008zk}.
 We also find complete agreement with the numerical result of Table~3 in 
\cite{Czakon:2008zk}, corresponding to a phase-space point in which the
$s,|t|,|u| \gg m^2$ approximation cannot be applied. With our code it is also
possible to reproduce the first plot in Figure~4 of \cite{Czakon:2008zk}, where
the  finite part of $A$ is shown as a surface depending on the variables $\eta$
and $\phi$, defined as
\be
\eta = \frac{s}{4 m^2} - 1 \, , \quad \phi = - \frac{t -m^2}{s} \,,
 \quad \frac{1}{2}\left(1-\sqrt{\frac{\eta}{1+\eta}} \right) \le \phi
 \le \frac{1}{2}\left(1+\sqrt{\frac{\eta}{1+\eta}} \right).
\ee
This surface is shown on the left side of Figure~\ref{surface}.

Furthermore, it is possible to expand our result for values of the center of mass
energy close to the production threshold. We define
\be
 \beta = \sqrt{1-\frac{4 m^2}{s}} \, , \qquad
 \xi = \frac{1-\cos\theta}{2} \, , \qquad
 L_\mu = \ln\left(\frac{\mu^2}{m^2} \right) \, , 
 \label{deff}
\ee
where $\theta$ is the scattering angle in the partonic center of mass frame,
and we expand our results in powers of the heavy quark velocity $\beta$, up to
terms of order $\beta^2$. The coefficients of this expansion contain
transcendental constants which originate from one- and two-dimensional HPLs 
evaluated at $x=1$. Since we did not find a satisfactory analytical
representation for all of these
constants, in the formulae below we present them in a 
numerical form.
We find:
\begin{eqnarray}
A(\beta,\xi)&=& \frac{A^{( -4)}(\beta,\xi)}{\vep^4} + \frac{A^{( -3)}(\beta,\xi)}{\vep^3} + \frac{A^{( -2)}(\beta,\xi)}{\vep^2} + \frac{A^{( -1)}(\beta,\xi)}{\vep} + A^{(0)}(\beta,\xi) + {\mathcal O}\left(\vep\right)\, , \nn \\
A^{( -4)} &=&
0.25 - 0.5 \beta^2 \left(1 -\xi \right) \xi+ {\mathcal O }\left(\beta^3\right)\, , \nn \\
A^{( -3)} &=&
1.68185 + 0.5 L_\mu + \beta (1 - 2 \xi) - \beta^2 \Bigl[ 
0.5 + \xi ( 1-\xi) (5.86371 + L_\mu) \Bigr]
+ {\mathcal O }\left(\beta^3\right) \, , \nn \\
A^{( -2)} &=&-2.67119 - 0.302961 L_\mu + 
 0.5 L_\mu^2 + \beta \Bigl[1.84475+ 
    2 L_\mu - \xi(3.68951 + 4 L_\mu) \Bigr]
\nn \\ & &    
     + \beta^2 \Bigl[0.777936- 
     L_\mu - \xi (1-\xi)(7.03784 + 4.39408 L_\mu +  L_\mu^2)  \Bigr]+{\mathcal O }\left(\beta^3\right) \, , \nn \\
A^{( -1)} &=&-8.15701 -  5.7593L_\mu - 2.13629 L_\mu^2 + 
 0.333333 L_\mu^3 + \beta \Bigl[-9.83935 - 3.64382 L_\mu
\nn \\ & & 
  + 
    2 L_\mu^2 + \xi\bigl(19.6787+ 7.28765 L_\mu - 
       4 L_\mu^2\bigr) \Bigr] + \beta^2 \Bigl[2.78693+ 5.22254 L_\mu 
\nn \\ & &       
       - 
    L_\mu^2 + \xi (1-\xi)(46.7006+  8.75816 L_\mu - 0.727411 L_\mu^2 - 
       0.666667 L_\mu^3)  \Bigr]+{\mathcal O }\left(\beta^3\right) \, ,  \nn \\
A^{(0)} &=& 23.5701 + 7.82592  L_\mu + 0.754463 L_\mu^2 - 2.03531 L_\mu^3 + 
 0.166667 L_\mu^4 + 
\nn \\ & & 
 \beta \Bigl[ 0.505501- 11.5953 L_\mu - 7.31049 L_\mu^2 + 
    1.33333 L_\mu^3 + \xi \bigl(-1.011+ 23.1906 L_\mu 
\nn \\ & &    
    + 14.621 L_\mu^2 - 
       2.66667 L_\mu^3\bigr) \Bigr] + \beta^2 \Bigl[-4.351 - 
       3.60348 L_\mu + 
    7.05587 L_\mu^2 
\nn \\ & &    
    - 
    0.666667 L_\mu^3 + \xi (1-\xi)\bigl(-5.36823 + 43.1864 L_\mu + 6.73063  L_\mu^2 + 
       0.737281 L_\mu^3
\nn \\ & &       
        - 0.333333 L_\mu^4\bigr)  \Bigr]+{\mathcal O }\left(\beta^3\right) \, .
\end{eqnarray}
Note that the dependence on $\beta$ and on $\xi$ in the formulae above is only 
polynomial. All the logarithmic terms 
$\ln{\beta}, \, \ln{\xi}, \, \ln{(1-\xi)}, \, \ln{(1-2\,\xi)}, \, \ldots$,
which are indeed present in the expansion of individual HPLs, cancel out in the final 
expressions. The coefficient $A$ is finite at threshold.
The expansion presented here could be used in the future for the calculation of
logarithmically enhanced terms near the $t \bar{t}$ production threshold. On the 
right hand side of Fig.~\ref{surface}, we compare the exact expression of the 
coefficient $A^{(0)}$ with the expansion in powers of $\beta$ including
up to terms of 
order $\beta^2$  (in the plot we set $\xi = 1/2$).

\section{Conclusions and Outlook\label{sec:conc}}

In this paper, we presented the analytic calculation of the two-loop planar
corrections to the heavy-quark production amplitude for $q\bar q \to t\bar t$,
retaining the exact heavy-quark mass dependence. The calculation of this subset
of diagrams allowed us to obtain an analytic expression for the leading color
coefficient in Eq.~(\ref{colstruc}). 
We checked our formula against recent results obtained in analytic form in
the small-mass region~\cite{Czakon:2007ej}. Moreover, we found numerical
agreement with the value of $A$ at the phase-space  point presented in
\cite{Czakon:2008zk}. 

Our result represents a gauge invariant sub-set of the full two-loop 
corrections to the partonic process $q \overline{q} \to t \overline{t}$. In
order to complete the analytic calculation  of the two-loop 
corrections, it is necessary to calculate the non-planar diagrams. Likewise,
analytic results for the two-loop amplitude for $gg\to t\bar t$ could be
obtained in the same calculational framework~\cite{ggtt}. However, some of the 
two-loop diagrams appearing in the gluon fusion channel cannot be expressed in terms 
of two-dimensional HPLs. In fact, their reduction to MIs  involves a
``sunrise''-type
subtopology with three equal massive propagators and an external momentum
which is not on the mass shell of the internal propagators. It is known that
already such a three-propagator graph involves elliptic integrals 
\cite{Laporta:2004rb}.

In order to obtain NNLO predictions for the total $t \overline{t}$ production
cross section and for differential distributions, it is necessary to combine the
two-loop virtual corrections with the already available~\cite{Dittmaier:2007wz}
one-loop corrections to the $t\bar t$+(1~parton) process and with the
tree-level  $t\bar t$+(2~partons) process.  These diagrams with additional
partons in the final state contribute to infrared-divergent configurations where
up to two partons can become unresolved. Their implementation
requires the application of a NNLO subtraction method. The methods presently
available~\cite{secdec,ant,cg} have been applied up to
now~\cite{babis,cghiggs,our3j,weinzierl3j}  to at most $1\to 3$ processes in
$e^+e^-$ annihilation and $2\to 1$ processes at hadron colliders. A  calculation
of a hadronic $2\to 2$ process, involving massive partons, will represent a new
step in complexity, potentially requiring further  refinements of the methods available to date.

\subsection*{Acknowledgments}

We would like to thank D.~Ma\^itre for his help and suggestions
and S.~Weinzierl for useful discussions. This research was supported in part
 by the Theory-LHC-France initiative of CNRS/IN2P3, 
by the Swiss National Science Foundation (SNF) 
under contract 200020-117602, and by the ``Forschungskredit der Universit\"at 
Z\"urich''.

\appendix

\section{Harmonic Polylogarithms \label{AppGPLs}}

The results presented here are conveniently expressed in terms of one- and
two-dimensional HPLs. Nowadays, harmonic polylogarithms are extensively
employed in multiloop computations. Therefore, in this appendix we only briefly
review their definition. The reader interested in the algebraic properties of
these functions can  find detailed discussions about this topic in the
available literature \cite{HPLs,Vollinga,2dHPLs,Aglietti:2004tq}.

In the non-physical region $s < 0$, seven  weight functions are needed for the
HPLs  with argument $x$. They are\footnote{
The last two weights in Eq.~(\ref{app1eq1}) introduce explicit 
imaginary parts in the formulae. 
However, these HPLs appear in such a way that these imaginary
parts cancel in the non-physical region, where the result has to be real.
Alternatively, one could choose a basis of weight functions in which the pair
$\frac{1}{2} \pm \frac{i \sqrt{3}}{2}$ is replaced by the original quadratic
expressions in the integrals: $1/(x^2-x+1)$ and $x/(x^2-x+1)$ 
\cite{Aglietti:2004tq}. In this case these HPLs are all manifestly real.}
\be
f_w(x) = \frac{1}{x-w}\, ,\quad \mbox{with} \quad w \in 
\left\{ 0,1,-1,-y,-\frac{1}{y},\frac{1}{2} \pm \frac{i \sqrt{3}}{2} \right\}
\, .
\label{app1eq1}
\ee
For HPLs with argument $y$, we need six weight functions
\be
f_w(y) =\frac{1}{y-w}\, ,\quad \mbox{with} \quad w \in 
\left\{ 0,1,-1,-x,-\frac{1}{x},1-\frac{1}{x} - x \right\} \, .
\ee
The weight-one HPLs are defined as
\be
G(0;x) = \ln{x} \, , \qquad G(w;x) = \int_0^x dt f_w(t) \, .
\ee
 HPLs of higher weight are defined by iterated  integrations
\be
G(w,\cdots;x) = \int_0^x dt f_w(t) G(\cdots;t) \, ,
\ee
with the only exception being the HPLs in which all the weights are zero which are
defined as follows
\be
G(\underbrace{0,0,\cdots,0}_n;x) = \frac{1}{n!} \ln^n{x} \,.
\ee
Analogous definitions hold for HPLs with argument $y$. The reader should be
aware of the fact that in the original definition of Remiddi and Vermaseren,
the weight function corresponding to the weight $1$ was $f_1 =1/(1-x)$. In
order to translate the HPLs defined with the Remiddi-Vermaseren convention to
the ones employed in this work (and vice versa) it is sufficient to multiply
each HPL by a factor $(-1)^n$, where $n$ is the number of weights equal to $1$.

The weights $-y$ and $-1/y$ were already introduced in
\cite{2dHPLs,electronloop,quarkloops}. In our results, the two-dimensional
harmonic polylogarithms have maximum weight four.   Therefore, it was not
possible to rewrite all the two-dimensional HPLs in terms of Nielsen 
polylogarithms, as it was done for the results of~\cite{quarkloops}, where  the
two-dimensional HPLs had maximal weight three. However, it is possible to
evaluate all the HPLs appearing in the analytic  expression of the coefficient
$A$ in Eq.~(\ref{colstruc}) by employing the  {\tt GiNaC} implementation of
multiple polylogarithms by Vollinga and  Weinzierl~\cite{Vollinga}.

We first obtained the squared matrix elements in the  non-physical region $s <
0$. The corresponding quantities in the  physical region $s > 4 m^2$  could be
obtained by analytic continuation to the complex value $s \to s + i \delta$,
where $\delta \to 0^+$. For $s > 4 m^2$ the variable $x$ becomes
\be
x = -x' + i \delta \, ,
\ee
where
\be
x' = \frac{\sqrt{s}-\sqrt{s-4 m^2}}{\sqrt{s} + \sqrt{s-4 m^2}} \, ,
\ee
So that $0<x'<1$ for $4 m^2<s<\infty$. The HPLs of argument $x$ can develop an 
imaginary part because of analytic continuation\footnote{
The coefficient $A$ is real for $s<0$. However, because of the weight functions
we use, and because of the fact that $0\le y \le \infty$, individual HPLs can
develop imaginary parts also in the non-physical region. The latters cancel out
among each other.}. In particular, the imaginary part of the HPLs of argument
$x$ for $s > 4 m^2$ is defined when the analytic continuation of the logarithm
is specified:
\be \label{ancon}
G(0;x) = G(0;-x'+ i \delta) = G(0,x') + i \pi\, .
\ee
For notation convenience, after the analytic continuation we rename $x'$ as $x$.

\section{Expansion of the HPLs near the Threshold}

We devote this section to a brief discussion of the technique employed to expand
the coefficient $A$ near the production threshold $\beta =\sqrt{1-4 m^2/s}=0$.

The first step consists in carrying out the analytic continuation from the
non-physical region, $s <0$, to the physical one, $s >4 m^2$, according to the
method outlined above.
The one- and two-dimensional HPLs appearing in the analytically continued
expression of the coefficient $A$ must then  be expanded in the  $\beta \to 0$
limit. While the threshold expansion of the ordinary HPLs does not lead to any 
particular difficulty, the expansion of the two-dimensional HPLs is indeed
more  delicate. The reason is that, in the latter case, the expansion
parameter  $\beta$ appears in both the argument and the weights of the HPLs.
In fact, in the physical region one finds:
\be
x = \frac{1-\beta}{1+\beta} \, , \qquad 
y = \frac{1-\beta}{1+\beta} + \frac{4\beta}{1-\beta^2} \xi \, .
\label{xofbeta}
\ee
Moreover, the coefficient $A$ depends on two-dimensional HPLs of maximal weight 
four; therefore it is very challenging to obtain explicit analytic expressions
of these functions in terms of logarithms and polylogarithms of complicated 
arguments, which can be subsequently expanded in $\beta \to 0$. 
Such an approach should be replaced by a more direct and algorithmic method. 

In the following we describe the method which allows to extract directly the
expansion of a given two-dimensional HPL of weight $n$, assuming that the
expansion of the HPLs of weight $n-1$ is known. Let us consider, for
simplicity, the case in which the HPL has argument $x$ and it has $y$ (or
$1/y$) in the  weights. Since the dependence of $x$ on $\beta$ is the one shown
in Eq.~(\ref{xofbeta}), we first use the transformation that relates 
$G(w,...;(1-\beta)/(1+\beta))$ to $G(w',...;\beta)$. This transformation
allows  to rewrite the HPL function of $x$ as a combination of HPL functions
of  $\beta$, HPLs with $y$ in the weights but evaluated in $x=1$, and HPLs
(either one- or two-dimensional) of a smaller weight.
The series expansion of the HPLs with argument $\beta$ is found recursively. 
We write them as the integral between 0 and $\beta$ of the total derivative
with  respect to $\beta$ of the HPLs themselves. The total derivative gives
rise to  HPLs of lower weight. We insert the expansions of the lower-weight
HPLs and  then we integrate again.
For the HPLs evaluated in $x=1$ the procedure is analogous.
It can happen that in the intermediate steps (expansion followed by an 
integration) logarithmic divergences occur. These divergences must be
regularized and they cancel in the final expressions.

Let us illustrate the algorithm more in detail. To this purpose, we consider
the example of a simple two-dimensional HPL of weight two, which appears in the
expression of  the coefficient $A$. 
For $0<x<1$ and $x<y<1/x$ we define
\be
G(y,-1;x) = \int_0^x dt \frac{1}{t-y} G(-1,t) = 
 \int_0^x dt \frac{1}{t-y} \ln(1+t) \, . 
 \label{ab1}
\ee
This HPL is real in the physical region, however the intermediate stages of
the  procedure described below require the sum of complex terms. We assume that
any ambiguity of this sort is dealt with by assigning an infinitesimal
imaginary part to $y$.
We start by rewriting the HPL of argument $x$ in terms of HPLs of argument
$\beta$; the integral representation in Eq.~(\ref{ab1}) can be rewritten as 
\be
G(y,-1;x) = G(y,-1;1) + \int_1^x  dt \frac{1}{t-y} G(-1,t) \, . 
\label{ab2}
\ee
In the integral in the second term of Eq.~(\ref{ab2}) we carry out a change of 
integration variable by setting $q = (1-t)/(1+t)$ ($dt = -2/(1+q)^2 dq$)
\bea
\int_1^x  dt \frac{1}{t-y} G(-1,t) &=& 
-2\int_0^\beta dq \frac{(1+q)}{(1-q) -y(1+q)} \frac{2}{(1+q)^2} 
G\left(-1,\frac{1-q}{1+q}\right) \, , \nn\\
&=& \int_0^\beta dq \, \left(\frac{1}{q-\gamma}- \frac{1}{q+1} \right)
G\left(-1;\frac{1-q}{1+q}\right) \, , \label{ab3}
\eea
where we employed the relation $\beta = (1-x)/(1+x)$ and we introduced $\gamma
= (1-y)/(1+y)$. We emphasize that, given the definitions of $x$ and $y$ in the
physical region, $\gamma$ is a function of $\beta$ and of the  variable $\xi$
defined in Eq.~(\ref{deff}). The weight one HPLs appearing in the integrand of
Eq.~(\ref{ab3}) can also be easily rewritten in terms of HPLs of argument $q$:
\be
G\left(-1;\frac{1-q}{1+q}\right) = -G(-1;q) + \ln{2} \, .
\ee
At this stage it is straightforward to integrate in $q$ to finally obtain
\bea
G(y,-1;x) &=& G(y,-1;1) - \Bigl( G(\gamma,-1;\beta) - G(-1,-1;\beta) \Bigr)
\nn \\
& &
+\ln{2} \Bigl( G(\gamma;\beta) -G(-1;\beta)\Bigr) \, . 
\label{ab7}
\eea
In the equation above, there are one- and two-dimensional HPLs of weight one,
that we consider to be known, and three HPLs of weight two. Among them, there
is  a single two-dimensional HPL of weight two, 
$G\left(\gamma(\beta),-1;\beta\right)$, and two one-dimensional HPLs. One of
them, $G(-1,-1;\beta)$, has a trivial expansion in $\beta \to 0$.  Let us
discuss the method employed to obtain the threshold expansion of the other two:
$G\left(\gamma(\beta),-1;\beta\right)$, and  $G(y,-1;1)$.

We can rewrite $G\left(\gamma(\beta),-1;\beta\right)$ as follows:
\be
G\left(\gamma(\beta),-1;\beta\right) = \int_0^\beta d\beta' \frac{d}{d \beta'}
G\left(\gamma(\beta'),-1;\beta'\right) + G\left(\gamma(0),-1;0\right) \, .
\label{ab5}
\ee
In this simple example the second term in Eq.~(\ref{ab5}) is well defined (and
it is actually equal to zero). The derivative in the first term of 
Eq.~(\ref{ab5}) can be rewritten as
\bea
\frac{d}{d \beta}
G\left(\gamma(\beta),-1;\beta\right) &=& \frac{1}{\beta - \gamma(\beta)}
G(-1;\beta) + \int_0^\beta dt \frac{d}{d \beta} \left(\frac{1}{t - \gamma(\beta)} G(-1; t)\right)  \nn \\
&=& \frac{1}{\beta - \gamma(\beta)}
G(-1;\beta) + \frac{d \gamma(\beta)}{d \beta} \int_0^\beta \frac{dt}{(t-\gamma(\beta))^2} G(-1;t) \nn \\
 &=& \frac{1}{\beta - \gamma(\beta)}
G(-1;\beta) - \frac{d \gamma(\beta)}{d \beta} \int_0^\beta dt \frac{d}{dt}\left(\frac{1}{t-\gamma(\beta)} \right) G(-1;t)\nn \\
 &=& \frac{1}{\beta - \gamma(\beta)}
G(-1;\beta) - \frac{d \gamma(\beta)}{d \beta} \Biggl[
\frac{1}{\beta -\gamma(\beta)} G(-1;\beta) \nn \\
& & 
-
 \frac{1}{1+\gamma(\beta)}\int_0^\beta dt \left(\frac{1}{t-\gamma(\beta)} - \frac{1}{t+1} \right) \Biggr]\nn \\ 
 &=& \frac{1}{\beta - \gamma}
G(-1;\beta) -\frac{d \gamma}{d \beta}\Biggl[ \frac{1}{\beta - \gamma}G(-1;\beta) -  \frac{1}{1+\gamma}\Bigl(
G(\gamma; \beta) \nn \\
& &-
 G(-1;\beta)\Bigr)\Biggr] \, ,
\eea
where in the last line we dropped the dependence of $\gamma$ on $\beta$. Since
the expansion of the HPLs of weight one is assumed to be known, it is 
straightforward to expand the equation above in the limit $\beta \to 0$ and  to
insert it in Eq.~(\ref{ab5}) to obtain
\be
G\left(\gamma,-1;\beta\right) = \beta \Bigl[ 1 + (1-2 \xi) \Bigl( \ln{2} + \ln{(\xi)}  - \ln{(2 \xi -1)}\Bigr)\Bigr] + {\mathcal O}(\beta^2) \, .
\ee
The formula above is real for $\xi > 1/2$. However, the imaginary parts which
arise for $\xi < 1/2$ cancel against the imaginary parts coming from the
expansion of $G(y,-1;1)$.

The expansion of $G(y,-1;1)$ can be done with the same algorithm. We write:
\be
G(y(\beta),-1;1) = \int_{0^+}^{\beta} \frac{d}{d \beta'} G(y(\beta'),-1;1)
+ G(y(0^+),-1;1) \, ,
\label{eqA17}
\ee
where $0^+$ indicates the fact that we must take the limit $\beta \to 0^+$
both in the integration constant and in the lower boundary of the integration.
Both limits are logarithmically divergent, but the divergencies cancel between
the two terms. We have:
\bea 
\frac{d}{d \beta'} G(y(\beta'),-1;1) & = & \frac{dy(\beta)}{d \beta} 
\int_0^1 \frac{dt}{(t-y)^2} G(-1,t) \, , \nn\\
& = & \frac{dy(\beta)}{d \beta} \left[ \left. - \frac{1}{t-y} G(-1,t) \right|
_0^1 + \frac{1}{1+y} \int_0^1 dt \left( \frac{1}{t-y} - \frac{1}{1+t} \right)  
\right] \, , \nn\\
& = & \frac{dy(\beta)}{d \beta} \left[ - \frac{\ln{2}}{1-y} + 
\frac{1}{1+y} \Bigl( G(y;1) - \ln{2} \Bigr) \right] \, .
\label{eqA18}
\eea

Expanding Eq.~(\ref{eqA18}) in the limit $\beta \to 0$, we find the following
Laurent series:
\be
\frac{d}{d \beta'} G(y(\beta'),-1;1) = \frac{\ln{2}}{\beta} 
+ \frac{\ln{2}}{2\xi-1} + (2\xi -1) \left[ \ln{(\beta)} 
+ \ln{(2\xi -1)} \right] + {\mathcal O}(\beta) \, .
\label{eqA19}
\ee

Integrating Eq.~(\ref{eqA19}), as in Eq.~(\ref{eqA17}), we find:
\bea
\int_{0^+}^{\beta} \frac{d}{d \beta'} G(y(\beta'),-1;1) & = & 
\ln{2}\ln{(\beta)}
- \ln{2}\ln{(0^+)} + \beta \biggl\{ \frac{\ln{2}}{2\xi-1} 
+ (2\xi -1) \Bigl[ \ln{(\beta)}  \nn\\
& &
+ \ln{(2\xi -1)} - 1 \Bigr] \biggr\}
+ {\mathcal O}(\beta^2)  .
\eea

Moreover, one finds:
\bea
G(y(0^+),-1;1) & = & G(y(0^+);1)G(-1;1) - G(-1,y(0^+);1) \, , \nn\\
& = & \ln{2} \left[ \ln{(0^+)} + \ln{2} + \ln{(2 \xi -1)} \right] 
+ \frac{1}{2} \zeta_2 - \frac{1}{2} \ln^2{2} \, .
\eea

Finally, we have:
\bea
G(y(\beta),-1;1) & = & \ln{2}\ln{(\beta)} + \ln{2} \ln{(2 \xi -1)} 
+ \frac{1}{2} \ln^2{2} + \frac{1}{2} \zeta_2  \nn\\
& &
+ \beta \biggl\{ \frac{\ln{2}}{2\xi-1} 
+ (2\xi -1) \Bigl[ \ln{(\beta)} 
+ \ln{(2\xi -1)} - 1 \Bigr] \biggr\}
+ {\mathcal O}(\beta^2)  \, ,
\eea
where the divergencies disappeared.
Considering also the expansion of the one- and two-dimensional HPLs of weight 1
and the expansion of $G(-1,-1;\beta)$ in Eq.~(\ref{ab7}), we find the final 
formula:
\bea
G(y,-1;x) & = & \ln{2}\ln{(\beta)} + \ln{2} \ln{(\xi)} 
+ \frac{3}{2} \ln^2{2} + \frac{1}{2} \zeta_2  \nn\\
& &
+ \beta \Bigl\{ - 2 \xi
+ (2\xi -1) \bigl[ \ln{(\beta)} 
+ \ln{(\xi)} + \ln{2} \bigr] \Bigr\}
+ {\mathcal O}(\beta^2)  \, .
\eea

In the case in which the argument of the HPLs is $y$ and $x$ is present in the 
weights, $G(w,...;y)$, the procedure is analogous to the one explained above. 
Since the dependence of $y$ on $\beta$ involves also the parameter $\xi$ 
(see Eq.~(\ref{xofbeta})), the first step consists in using the scale properties 
of the HPLs:
\be
G(w,...;y) = G(\lambda w,...;\lambda y)\, ,
\ee
(valid in the case in which all the trailing zeroes have already been extracted)
to get rid of the $\xi$ dependence in the argument. To achieve this goal, one multiplies
weights and argument by
\be
\lambda = \frac{1}{1 + \frac{4 \beta}{(1-\beta)^2} \xi} \, .
\ee
In so doing, we fit again in the case illustrated in the example concerning
$G(w,...;x)$, since
\be
\frac{1}{1 + \frac{4 \beta}{(1-\beta)^2} \xi} \, y = \frac{1-\beta}{1+\beta} \, ,
\ee
and the expansion proceeds along the same steps as outlined above.

\section{Master Integrals \label{AppMIs}}

In this Appendix we collect the MIs for the topologies in Fig.~\ref{nlMIs}.

The explicit expression of the MIs depends on the chosen normalization of the
integration measure. The integration on the loop momenta is normalized as
follows:
\be
\int{\mathfrak
D}^dk = \frac{1}{C(\vep)} \left( \frac{\mu^2}{m^2}
\right)^{-\vep} \mu^{2 \vep} \! \int \frac{d^d k}{(4 \pi^2)^{(1-\vep)}} \, ,
\label{measure}
\ee
where $C(\vep)$ was defined in Eq.~(\ref{Cep}). In Eq.~(\ref{measure}) $\mu$
stands for the 't Hooft mass of dimensional regularization. The integration
measure in Eq.~(\ref{measure}) is chosen in such a way that the one-loop
massive tadpole becomes 
\be
\int{\mathfrak D}^dk \ \frac{1}{k^2+m^2} =
              -\frac {m^2}{4 (1- \vep) \vep} \, .
\label{Tadpole}
\ee
In calculating the squared matrix element, we multiply our bare results by
$(\mu^2/m^2)^\vep$, in order to make explicit the dependence on the 't Hooft
scale. We also point out that, since the squared matrix element still contains
soft and collinear divergencies regulated by $\ep$, it depends on the
normalization of the integration measure. In particular, in order to match our
results with the ones of \cite{Czakon:2007ej,Czakon:2008zk}, it is necessary to
multiply the latter by the factor
\be
\frac{e^{-2 \gamma_{\mbox{{\tiny}}} \vep}}{\Gamma\left(1+\vep \right)^2} =
1- \zeta(2) \vep^2 + \frac{2}{3} \zeta(3) \vep^3 + \frac{3}{10} \zeta(2)^2 
\vep^4 + {\mathcal O}\left( \vep^5\right) \, .
\ee
The MIs are expanded in powers of the dimensional regulator $\vep$; below we
collect the analytic expression of the coefficients in the $\vep$ expansion up
to terms involving HPLs and 2dHPLs of weight three. The coefficients
involving HPLs and 2dHPLs of weight four are also needed in order to obtain the
finite part of the leading color coefficient, and we calculated them. However,
their analytic expressions are too long to be written in this appendix; the
interested reader can find them in the text file included in the arXiv
submission of this paper. 

There are two MIs belonging to the topology Fig.~\ref{nlMIs}-(a).
The first MI is

\begin{eqnarray}
\hbox{\begin{picture}(0,0)(0,0)
\SetScale{1}
  \SetWidth{.5}
  \Line(-30,20)(30,20)
  \Line(-30,-20)(30,-20)
  \Line(-20,20)(-20,-20)
  \Line(-20,20)(20,20)
\CArc(20,0)(20,90,270)
  \SetWidth{1.6}
  \Line(20,20)(30,20)
  \Line(20,-20)(30,-20)
  \Line(20,20)(20,-20)
%
\end{picture}}
& \hspace*{10mm} = & \int \frac{{\mathfrak
D}^dk_1  {\mathfrak
D}^dk_2}{P_0\left(k_2 \right) P_0\left(k_1-k_2\right)
P_0 \left(k_2-p_1\right) P_0\left(k_2-p_1-p_2\right) P_m\left(k_1-p_3\right)} \, ,
\end{eqnarray}

\vspace*{5mm}
\noindent where we define
\be
P_0\left(k\right) \equiv k^2 \, , \qquad P_m\left(k\right) \equiv k^2 +m^2 \, .
\ee
We find

\begin{eqnarray}
\hbox{\begin{picture}(0,0)(0,0)
\SetScale{1}
  \SetWidth{.5}
  \Line(-30,20)(30,20)
  \Line(-30,-20)(30,-20)
  \Line(-20,20)(-20,-20)
  \Line(-20,20)(20,20)
\CArc(20,0)(20,90,270)
  \SetWidth{1.6}
  \Line(20,20)(30,20)
  \Line(20,-20)(30,-20)
  \Line(20,20)(20,-20)
%
\end{picture}}
& \hspace*{16mm} = &\frac{1}{m^2}  \sum_{i=-3}^0 A_i \vep^i  + {\mathcal O}(\vep) \, ,
\end{eqnarray}

\begin{eqnarray} \nonumber
 A_{-3} &=&  \frac{x }{16 (1-x)^2  }\, , \nn \\
 A_{-2} &=& -\frac{x}{16 (1-x)^2 y} \Bigl[ -2  y-  y G(0;x) +2  y G(1;x) 
            + (y+1) G(-1;y)\Bigr]\, , \nn \\
 A_{-1} &=& -\frac{x}{16 (1-x)^2 y} \Bigl[  y (3 \zeta(2)-4)+2  (y+1) G(-1;y)-2  y G(0;x)
\nn \\
& &+  (y+1)
G(-1;y) G(0;x)+ 4  y G(1;x)-2  (y+1) G(-1;y) G(1;x)
\nn \\
& &-2  (y+1) 
G(-1,-1;y)+  (y-1) G(0,-1;y)-  y G(0,0;x)+2  y G(0,1;x)
\nn \\
&&+2  y 
G(1,0;x)-4  y G(1,1;x)\Bigr]  \, , \nn \\
A_{0} &=& -\frac{x}{16 (1-x)^2 y} \Bigl[ 2  y (3 \zeta(2)+7 \zeta(3)-4)- (7 \zeta(2) y-4 y+11
\zeta(2)-4) G(-1;y)
\nn \\
& &- y (\zeta(2)+4) G(0;x)+2  (y+1) G(-1;y) 
G(0;x)-2  y (3 \zeta(2)-4) G(1;x)
\nn \\
& &-4  (y+1) G(-1;y) G(1;x)+4  
(y+1) \zeta(2) G\left(-1/y;x\right)-4  (y+1) G(-1,-1;y)
\nn \\
& &-4  
(y+1) G(0;x) G(-1,-1;y)+4  (y+1) G(1;x) G(-1,-1;y)
\nn \\
& &+2  (y+1) 
G\left(-1/y;x\right) G(-1,-1;y) +2  (y+1) G(-y;x) 
G(-1,-1;y)
\nn \\
& &+2  (y-1) G(0,-1;y) +
2  y G(0;x) G(0,-1;y)-2  (y-1) 
G(1;x) G(0,-1;y)
\nn \\
& &- (y+1) G\left(-1/y;x\right) G(0,-1;y)- 
(y+1) G(-y;x) G(0,-1;y)-2  y G(0,0;x)
\nn \\
& &+4  y G(0,1;x)+4  y 
G(1,0;x)-2  (y+1) G(-1;y) G(1,0;x)-8  y G(1,1;x)
\nn \\
& &+4  (y+1) G(-1;y) 
G(1,1;x)+ (y+1) G(-1;y) G\left(-1/y,0;x\right)
\nn \\
& &-2  (y+1) 
G(-1;y) G\left(-1/y,1;x\right)+ (y+1) G(-1;y) G(-y,0;x)
\nn \\
& &-2  
(y+1) G(-1;y) G(-y,1;x)+2  (y-1) G(-1,0,-1;y)+4  G(0,-1,-1;y)
\nn \\
& &-2  y 
G(0,0,-1;y)-2  y G(0,0,0;x)+4  y G(0,0,1;x)+ (3 y+1) G(0,1,0;x)
\nn \\
& &-2 
 (3 y+1) G(0,1,1;x)+2  y G(1,0,0;x)-4  y G(1,0,1;x)-4  y 
G(1,1,0;x)
\nn \\
& &
+8  y G(1,1,1;x)+ (y+1) G\left(-1/y,0,0,x
\right)-2  (y+1) G\left(-1/y,0,1;x\right)
\nn \\
& &- (y+1) 
G\left(-1/y,1,0;x\right)+2  (y+1) G\left(-1/y,1,1,x
\right)- (y+1) G(-y,1,0;x)
\nn \\
& &+2  (y+1) G(-y,1,1;x) \Bigr] \, .
\end{eqnarray}

The second MI for the topology in Fig.~\ref{nlMIs}-(a) was chosen as

\begin{eqnarray}
\hbox{\begin{picture}(0,0)(0,0)
\SetScale{1}
  \SetWidth{.5}
  \Line(-30,20)(30,20)
  \Line(-30,-20)(30,-20)
  \Line(-20,20)(-20,-20)
  \Line(-20,20)(20,20)
\CArc(20,0)(20,90,270)
  \SetWidth{1.6}
  \GCirc(20,0){3}{0}
  \Line(20,20)(30,20)
  \Line(20,-20)(30,-20)
  \Line(20,20)(20,-20)
%
\end{picture}}
& \hspace*{10mm} = & \int \frac{{\mathfrak
D}^dk_1  {\mathfrak
D}^dk_2}{P_0\left(k_2 \right) P_0\left(k_1-k_2\right)
P_0 \left(k_2-p_1\right) P_0\left(k_2-p_1-p_2\right) P^2_m\left(k_1-p_3\right)} \, .
\end{eqnarray}

\vspace*{5mm}
Its analytic expression is

\begin{eqnarray}
\hbox{\begin{picture}(0,0)(0,0)
\SetScale{1}
  \SetWidth{.5}
  \Line(-30,20)(30,20)
  \Line(-30,-20)(30,-20)
  \Line(-20,20)(-20,-20)
  \Line(-20,20)(20,20)
\CArc(20,0)(20,90,270)
  \SetWidth{1.6}
  \GCirc(20,0){3}{0}
  \Line(20,20)(30,20)
  \Line(20,-20)(30,-20)
  \Line(20,20)(20,-20)
%
\end{picture}}
& \hspace*{16mm} = &\frac{1}{m^4} \sum_{i=-2}^0 A_i \vep^i  + {\mathcal O}(\vep) \, ,
\end{eqnarray}

\begin{eqnarray} 
 A_{-2} &=& \frac{x}{16 (1-x)^2 y} G(-1;y)\, , \nn \\
 A_{-1} &=& -\frac{x}{16 (1-x)^2 y} \Bigl[ - G(-1;y) G(0;x)+2  G(-1;y) G(1;x)+2  G(-1,-1;y)\nn \\
 &&+ G(0,-1;y)\Bigr]\, , \nn \\
  A_{0} &=& -\frac{x}{16 (1-x)^2 y} \Bigl[11  \zeta(2) G(-1;y)+2  G(1,0;x) G(-1;y)-4  G(1,1;x) G(-1;y)
\nn \\  
& & -  
G\left(-1/y,0;x\right) G(-1;y)+2  G\left(-1/y,1;x\right) G(-1;y)- G(-y,0;x) G(-1;y)
\nn \\
& &
+2  G(-y,1;x) G(-1;y)-4  \zeta(2) \
G\left(-1/y;x\right)+4  G(0;x) G(-1,-1;y)
\nn \\
& & -4  G(1;x) 
G(-1,-1;y)-2  G\left(-1/y;x\right) G(-1,-1;y)-2  G(-y;x) 
G(-1,-1;y)
\nn \\
& &
-2  G(1;x) G(0,-1;y)+ G\left(-1/y;x\right) 
G(0,-1;y)+ G(-y;x) G(0,-1;y)
\nn \\
& &
+2  G(-1,0,-1;y)-4  G(0,-1,-1;y)- 
G(0,1,0;x)+2  G(0,1,1;x)
\nn \\
& &- G\left(-1/y,0,0;x\right)\!+\!2  
G\left(-1/y,0,1;x\right)\!+\! G\left(-1/y,1,0;x\right)
\nn \\
&& -2 
 G\left(-1/y,1,1;x\right)+ G(-y,1,0;x)-2  G(-y,1,1;x)\Bigr] \, .
\end{eqnarray}

The topology shown in Fig.~\ref{nlMIs}-(b) has two MIs. The first one,

\begin{eqnarray}
\hbox{\begin{picture}(0,0)(0,0)
\SetScale{1}
  \SetWidth{.5}
  \Line(-30,20)(30,20)
  \Line(-30,-20)(30,-20)
  \Line(-20,20)(-20,-20)
  \Line(-20,20)(20,20)
\CArc(-20,0)(20,270,450)
  \SetWidth{1.6}
  \Line(20,20)(30,20)
  \Line(20,-20)(30,-20)
  \Line(20,20)(20,-20)
\end{picture}}
& \hspace*{10mm} = & \int \frac{{\mathfrak
D}^dk_1  {\mathfrak
D}^dk_2}{P_0\left(k_1 \right) P_0\left(k_1-k_2\right)
P_0 \left(k_2-p_1\right) P_0\left(k_1-p_1-p_2\right) P_m\left(k_1-p_3\right)} 
\, ,
\end{eqnarray}

\vspace*{5mm}
\noindent has the following analytic expression:

\begin{eqnarray}
\hbox{\begin{picture}(0,0)(0,0)
\SetScale{1}
  \SetWidth{.5}
  \Line(-30,20)(30,20)
  \Line(-30,-20)(30,-20)
  \Line(-20,20)(-20,-20)
  \Line(-20,20)(20,20)
\CArc(-20,0)(20,270,450)
  \SetWidth{1.6}
  \Line(20,20)(30,20)
  \Line(20,-20)(30,-20)
  \Line(20,20)(20,-20)
\end{picture}}
& \hspace*{16mm} = &\frac{1}{m^2} \sum_{i=-1}^0 A_i \vep^i  + {\mathcal O}(\vep) \, ,
\end{eqnarray}

\begin{eqnarray} 
 A_{-1} &=& \frac{x}{16 (1-x^2)} \left[ 4 \zeta(2)+G(0,0;x)-2 G(0,1;x)  \right]\, , \nn \\
 A_{0} &=& \frac{x}{16 (1-x^2)} \Bigl[\zeta(2)\Bigl( 16 G(-1;x) -G(0;x) -12 G(1;x) -4 \
G\left(-1/y;x\right)+8 \Bigr)
\nn \\ 
& &+7 \zeta(3)-2 \
G\left(-1/y;x\right) G(-1,-1;y)+2 G(-y;x) G(-1,-1;y)
\nn \\
& & +G(0;x) 
G(0,-1;y)+G\left(-1/y;x\right) G(0,-1;y)-G(-y;x) G(0,-1;y)+2 \
G(0,0;x)
\nn \\
& &-4 G(0,1;x)-G(-1;y) G\left(-1/y,0;x\right)+2 G(-1;y) \
G\left(-1/y,1;x\right)
\nn \\
& &+G(-1;y) G(-y,0;x)-2 G(-1;y) \
G(-y,1;x)+4 G(-1,0,0;x)-8 G(-1,0,1;x)
\nn \\
& &+2 G(0,-1,-1;y)-G(0,0,-1;y)+2 \
G(0,0,0;x)-4 G(0,0,1;x)-4 G(0,1,0;x)
\nn \\
& &+8 G(0,1,1;x)-3 G(1,0,0;x)+6 \
G(1,0,1;x)-G\left(-1/y,0,0;x\right)
\nn \\
& &+2 
G\left(-1/y,0,1;x\right)+G\left(-1/y,1,0;x\right)-2 G\
\left(-1/y,1,1;x\right)-G(-y,1,0;x)
\nn \\
& &+2 G(-y,1,1;x)\Bigr]\, . 
 \end{eqnarray}

The second MI for the topology in Fig.~\ref{nlMIs}-(b) is:

\begin{eqnarray}
\hbox{\begin{picture}(0,0)(0,0)
\SetScale{1}
  \SetWidth{.5}
  \Line(-30,20)(30,20)
  \Line(-30,-20)(30,-20)
  \Line(-20,20)(-20,-20)
  \Line(-20,20)(20,20)
\CArc(-20,0)(20,270,450)
  \SetWidth{1.6}
  \GCirc(-20,0){3}{0}
  \Line(20,20)(30,20)
  \Line(20,-20)(30,-20)
  \Line(20,20)(20,-20)
\end{picture}}
& \hspace*{10mm} = & \int \frac{{\mathfrak
D}^dk_1  {\mathfrak
D}^dk_2}{P_0\left(k_1 \right) P^2_0\left(k_1-k_2\right)
P_0 \left(k_2-p_1\right) P_0\left(k_1-p_1-p_2\right) P_m\left(k_1-p_3\right)} \, ,
\end{eqnarray}

\vspace*{5mm}
\noindent where 

 \begin{eqnarray}
\hbox{\begin{picture}(0,0)(0,0)
\SetScale{1}
  \SetWidth{.5}
  \Line(-30,20)(30,20)
  \Line(-30,-20)(30,-20)
  \Line(-20,20)(-20,-20)
  \Line(-20,20)(20,20)
\CArc(-20,0)(20,270,450)
  \SetWidth{1.6}
  \GCirc(-20,0){3}{0}
  \Line(20,20)(30,20)
  \Line(20,-20)(30,-20)
  \Line(20,20)(20,-20)
\end{picture}}
& \hspace*{16mm} = &\frac{1}{m^4} \sum_{i=-3}^0 A_i \vep^i  + {\mathcal O}(\vep) \, ,
\end{eqnarray}

\begin{eqnarray} 
 A_{-3} &=& -\frac{7 x}{192 (1-x)^2 (1+y)} \, ,  \nn \\
 A_{-2} &=& -\frac{ x}{96 (1-x)^2 (1+y)} \Bigl[-8 G(-1;y)+3 G(0;x)-6 G(1;x)  \Bigr] \, ,  \nn \\ 
 A_{-1} &=&-\frac{ x}{24 (1-x)^2 (1+y)} \Bigl[ -5 \zeta(2)-3 G(-1;y) G(0;x)+6 G(-1;y) G(1;x)
 \nn \\
 & &+2 G(-1,-1;y)\Bigr]\, ,  \nn \\ 
 A_{0} &=&-\frac{ x}{48 (1-x)^2 (1+y)} \Bigl[-29 \zeta(3)+46 \zeta(2) G(-1;y)+3 \zeta(2) G(0;x)
 \nn \\
 & &+30 
\zeta(2) G(1;x)-36 \zeta(2) G\left(-1/y;x\right)+24 G(0;x) 
G(-1,-1;y)
\nn \\ 
& &-12 G(1;x) G(-1,-1;y)-18 G\left(-1/y;x\right) 
G(-1,-1;y)-18 G(-y;x) G(-1,-1;y)
\nn \\
& &-9 G(0;x) G(0,-1;y)+9 
G\left(-1/y;x\right) G(0,-1;y)+9 G(-y;x) G(0,-1;y)
\nn \\
& &+18 
G(-1;y) G(1,0;x)-36 G(-1;y) G(1,1;x)-9 G(-1;y) 
G\left(-1/y,0;x\right)
\nn \\
& &+18 G(-1;y) G\left(-1/y,1;x\right)-9 G(-1;y) G(-y,0;x)+18 G(-1;y) G(-y,1;x)
\nn \\
& &+32 G(-1,-1,-1;y)-18 
G(-1,0,-1;y)-18 G(0,-1,-1;y)+9 G(0,0,-1;y)
\nn \\
& &+9 G(1,0,0;x)-18 
G(1,0,1;x)-18 G(1,1,0;x)+36 G(1,1,1;x)
\nn \\
& &-9 G\left(-1/y,0,0;x\right)+18 G\left(-1/y,0,1;x\right)+9
G\left(-1/y,1,0;x\right)
\nn \\
& &-18 G\left(-1/y,1,1;x\right)
+9 G(-y,1,0;x)-18 G(-y,1,1;x) \Bigr] \, .
\end{eqnarray} 
 
The are two MIs for the topology  in Fig.~\ref{nlMIs}-(c). The first one is

\begin{eqnarray}
\hbox{\begin{picture}(0,0)(0,0)
\SetScale{1}
  \SetWidth{.5}
  \Line(-30,20)(30,20)
  \Line(-30,-20)(30,-20)
  \Line(-20,20)(-20,-20)
  \Line(-20,20)(20,20)
  \Line(-20,-20)(20,20)
  \SetWidth{1.6}
  \Line(20,20)(30,20)
  \Line(20,-20)(30,-20)
  \Line(20,20)(20,-20)
\end{picture}}
& \hspace*{10mm} = & \int \frac{{\mathfrak
D}^dk_1  {\mathfrak
D}^dk_2}{P_0\left(k_2 \right) P_0\left(k_1-k_2\right)
P_0 \left(k_2-p_1\right) P_0\left(k_1-p_1-p_2\right) P_m\left(k_1-p_3\right)} \, ,
\end{eqnarray}

\vspace*{5mm}
\noindent and the corresponding Laurent expansion in $\vep$ is 

 \begin{eqnarray}
\hbox{\begin{picture}(0,0)(0,0)
\SetScale{1}
  \SetWidth{.5}
  \Line(-30,20)(30,20)
  \Line(-30,-20)(30,-20)
  \Line(-20,20)(-20,-20)
  \Line(-20,20)(20,20)
  \Line(-20,-20)(20,20)
  \SetWidth{1.6}
  \Line(20,20)(30,20)
  \Line(20,-20)(30,-20)
  \Line(20,20)(20,-20)
\end{picture}}
& \hspace*{16mm} = &\frac{1}{m^2}  \frac{A_{-1}}{\vep}   + {\mathcal O}(\vep^0) \, ,
\end{eqnarray}

\begin{eqnarray} 
 A_{-1} &=& - \frac{x}{16 (1-x+x^2+x y)}\Bigl[ \zeta(3)+ \zeta(2) G(-1;y)-3 \zeta(2) G(0;x)
 \nn \\
 & &-6 
\zeta(2) G(1;x)+4 \zeta(2) G\left(-1/y;x\right)-4 
G(1;x) G(-1,-1;y)
\nn \\
& &+2 G\left(-1/y;x\right) G(-1,-1;y)+2  
G(-y;x) G(-1,-1;y)+  G(0;x) G(0,-1;y)
\nn \\
& &-  G\left(-1/y;x\right) G(0,-1;y)-  G(-y;x) G(0,-1;y)-2  G(-1;y) G(1,0;x)
\nn \\
& &+4  
G(-1;y) G(1,1;x)+  G(-1;y) G\left(-1/y,0;x\right)-2
G(-1;y) G\left(-1/y,1;x\right)
\nn \\
& &+ G(-1;y) G(-y,0;x)-2 
G(-1;y) G(-y,1;x)+ G(-1,0,-1;y)
\nn \\
& &+2 G(0,-1,-1;y)- 
G(0,0,-1;y)- G(0,0,0;x)+2 G(0,0,1;x)
\nn \\
& & 
- G(1,0,0;x)+2 
G(1,0,1;x)+2 G(1,1,0;x)-4 G(1,1,1;x)
\nn \\
& &+ 
G\left(-1/y,0,0;x\right)-2 G\left(-1/y,0,1;x\right)- G\left(-1/y,1,0;x\right)
\nn \\
& &+2 
G\left(-1/y,1,1;x\right)- G(-y,1,0;x)+2 G(-y,1,1;x)\Bigr] \, . 
\end{eqnarray} 

The second MI for the topology in Fig.~\ref{nlMIs}-(c) was chosen to be

\begin{eqnarray}
\hbox{\begin{picture}(0,0)(0,0)
\SetScale{1}
  \SetWidth{.5}
  \Line(-30,20)(30,20)
  \Line(-30,-20)(30,-20)
  \Line(-20,20)(-20,-20)
  \Line(-20,20)(20,20)
  \Line(-20,-20)(20,20)
  \SetWidth{1.6}
  \GCirc(20,0){3}{0}
  \Line(20,20)(30,20)
  \Line(20,-20)(30,-20)
  \Line(20,20)(20,-20)
\end{picture}}
& \hspace*{10mm} = & \int \frac{{\mathfrak
D}^dk_1  {\mathfrak
D}^dk_2}{P_0\left(k_2 \right) P_0\left(k_1-k_2\right)
P_0 \left(k_2-p_1\right) P_0\left(k_1-p_1-p_2\right) P^2_m\left(k_1-p_3\right)} \, ,
\end{eqnarray} 

\vspace*{5mm}
\noindent where 

\begin{eqnarray}
\hbox{\begin{picture}(0,0)(0,0)
\SetScale{1}
  \SetWidth{.5}
  \Line(-30,20)(30,20)
  \Line(-30,-20)(30,-20)
  \Line(-20,20)(-20,-20)
  \Line(-20,20)(20,20)
  \Line(-20,-20)(20,20)
  \SetWidth{1.6}
  \GCirc(20,0){3}{0}
  \Line(20,20)(30,20)
  \Line(20,-20)(30,-20)
  \Line(20,20)(20,-20)
\end{picture}}
& \hspace*{16mm} = &\frac{1}{m^4} \sum_{i=-3}^0 A_{i} \vep^i   + {\mathcal O}(\vep) \, ,
\end{eqnarray}

\begin{eqnarray} 
 A_{-3} &=& -\frac{x}{96 (1-x)^2} \, , \nn \\
 A_{-2} &=&\frac{x}{96 (1-x)^2} \Bigl[ -2  G(-1;y)-3  G(0;x)+6  G(1;x) \Bigr] \, , \nn \\
 A_{-1} &=&\frac{x}{96 (1-x)^2} \Bigl[ -11  \zeta(2)-6  G(-1;y) G(0;x)+12  G(-1;y) G(1;x)-4  
G(-1,-1;y)
\nn \\
& &-9  G(0,0;x)+18  G(0,1;x)+18  G(1,0;x)-36  G(1,1;x) \Bigr] \, , \nn \\
 A_{-0} &=&\frac{x}{96 (1-x)^2} \Bigl[-34  \zeta(3)+50  \zeta(2) G(-1;y)+39  \zeta(2) G(0;x)+66  
\zeta(2) G(1;x)
 \nn \\
 & &-72  \zeta(2) G\left(-1/y;x\right)+24  
G(0;x) G(-1,-1;y)+24  G(1;x) G(-1,-1;y)
 \nn \\
 & &-36  G\left(-1/y;x\right) G(-1,-1;y)-36  G(-y;x) G(-1,-1;y)-18  G(0;x) G(0,-1;y)
 \nn \\
 & &+18  G
\left(-1/y;x\right) G(0,-1;y)+18  G(-y;x) G(0,-1;y)+36  
G(-1;y) G(1,0;x)
 \nn \\
 & &-72  G(-1;y) G(1,1;x)-18  G(-1;y) 
G\left(-1/y,0;x\right)+36  G(-1;y) G\left(-1/y,1;x\right)
 \nn \\
 & &-18  G(-1;y) G(-y,0;x)+36  G(-1;y) G(-y,1;x)+64  
G(-1,-1,-1;y)
 \nn \\
 & &-36  G(-1,0,-1;y)-36  G(0,-1,-1;y)+18  G(0,0,-1;y)-9 
 G(0,0,0;x)
 \nn \\
 & &+18  G(0,0,1;x)+36  G(0,1,0;x)-72  G(0,1,1;x)+54 
G(1,0,0;x)
 \nn \\
 & &-108  G(1,0,1;x)-108  G(1,1,0;x)+216  G(1,1,1;x)-18 
G\left(-1/y,0,0;x\right)
 \nn \\
 & &+36  G\left(-1/y,0,1;x\right)+18  G\left(-1/y,1,0;x\right)-36 
G\left(-1/y,1,1;x\right)
 \nn \\
 & &+18  G(-y,1,0;x)-36  G(-y,1,1;x) \Bigr] \, .
 \end{eqnarray} \\

The topology shown in Fig.~\ref{nlMIs}-(d) has two MIs. The first one is

\begin{eqnarray}
\hbox{\begin{picture}(0,0)(0,0)
\SetScale{1}
  \SetWidth{.5}
  \Line(-30,20)(30,20)
  \Line(-30,-20)(30,-20)
  \Line(-20,20)(-20,-20)
  \Line(-20,20)(20,20)
  \Line(-20,20)(20,0)
  \SetWidth{1.6}
  \Line(20,20)(30,20)
  \Line(20,-20)(30,-20)
  \Line(20,20)(20,-20)
\end{picture}}
& \hspace*{10mm} \!\!=\!\! & \int\! \!\! \frac{{\mathfrak
D}^dk_1  {\mathfrak
D}^dk_2}{P_0\!\left(k_2 \!\right) P_0\!\left(k_1\!-\!k_2\!\right)
P_0\!\left(k_1\!-\!p_1\!\right) P_0\!\left(k_1\!-\!p_1\!-\!p_2\!\right) 
P_m\!\left(k_1\!-\!p_3\!\right) P_m\!\left(k_2\!-\!p_3\!\right)} \, ,
\end{eqnarray} 

\vspace*{5mm}
\noindent with

\begin{eqnarray}
\hbox{\begin{picture}(0,0)(0,0)
\SetScale{1}
  \SetWidth{.5}
  \Line(-30,20)(30,20)
  \Line(-30,-20)(30,-20)
  \Line(-20,20)(-20,-20)
  \Line(-20,20)(20,20)
  \Line(-20,20)(20,0)
  \SetWidth{1.6}
  \Line(20,20)(30,20)
  \Line(20,-20)(30,-20)
  \Line(20,20)(20,-20)
\end{picture}}
& \hspace*{16mm} = &\frac{1}{m^4} \sum_{i=-2}^{-1} A_{i} \vep^i   + {\mathcal O}(\vep^0) \, ,
\end{eqnarray}

\begin{eqnarray} 
 A_{-2} &=& -\frac{x}{32 (1-x^2) (1+y)} \Bigl[-4  \zeta(2)
 - G(0,0;x)+2  G(0,1;x) \Bigr] \, , \nn \\
  A_{-1} &=& -\frac{x}{32 (1-x^2) (1+y)} \Bigl[
  -5  \zeta(3)+8  \zeta(2) G(-1;x)+8  \zeta(2) G(-1;y)-3  
\zeta(2) G(0;x)
\nn \\
& &+16  \zeta(2) G(1;x)-8  \zeta(2) \
G\left(-1/y;x\right)-4  G\left(-1/y;x\right) \
G(-1,-1;y)
\nn \\
& &
+4  G(-y;x) G(-1,-1;y)+2  G(0;x) G(0,-1;y)+2 
G\left(-1/y;x\right) G(0,-1;y)
\nn \\
& &
-2  G(-y;x) G(0,-1;y)+2 
G(-1;y) G(0,0;x)-4  G(-1;y) G(0,1;x)
\nn \\
& &
-2  G(-1;y) \
G\left(-1/y,0;x\right)+4  G(-1;y) G\left(-1/y,1;x\right)+2  G(-1;y) G(-y,0;x)
\nn  \\ 
& &-4  G(-1;y) G(-y,1;x)+2  G(-1,0,0;x)-4 
G(-1,0,1;x)+4  G(0,-1,-1;y)
\nn \\
& &
-2  G(0,0,-1;y)-3 G(0,0,0;x)+6 
G(0,0,1;x)+2  G(0,1,0;x)
\nn \\
& &
-4  G(0,1,1;x)+4  G(1,0,0;x)-8 
G(1,0,1;x)-2  G\left(-1/y,0,0;x\right)
\nn \\
& &
+4  
G\left(-1/y,0,1;x\right)+2  G\left(-1/y,1,0;x\right)-4  G\left(-1/y,1,1;x\right)-2  G(-y,1,0;x)
\nn \\ 
& &+4  
G(-y,1,1;x)
  \Bigr] \, .
\end{eqnarray} 

The second MI for topology ~\ref{nlMIs}-(d) is 

\begin{eqnarray}
\hbox{\begin{picture}(0,0)(0,0)
\SetScale{1}
  \SetWidth{.5}
  \Line(-30,20)(30,20)
  \Line(-30,-20)(30,-20)
  \Line(-20,20)(-20,-20)
  \Line(-20,20)(20,20)
  \Line(-20,20)(20,0)
\Text(30,-3)[cb]{$P_7$}
  \SetWidth{1.6}
  \Line(20,20)(30,20)
  \Line(20,-20)(30,-20)
  \Line(20,20)(20,-20)
\end{picture}}
& \hspace*{13mm} \!\!=\!\! & \int\! \!\! \frac{{\mathfrak
D}^dk_1  {\mathfrak
D}^dk_2 (k_2 - p_1 -p_2)^2}{P_0\!\left(k_2 \!\right) P_0\!\left(k_1\!-\!k_2\!\right)
P_0\!\left(k_1\!-\!p_1\!\right) P_0\!\left(k_1\!-\!p_1\!-\!p_2\!\right) 
P_m\!\left(k_1\!-\!p_3\!\right) P_m\!\left(k_2\!-\!p_3\!\right)} \, ,
\end{eqnarray} 

\vspace*{5mm}
\noindent where

\begin{eqnarray}
\hbox{\begin{picture}(0,0)(0,0)
\SetScale{1}
  \SetWidth{.5}
  \Line(-30,20)(30,20)
  \Line(-30,-20)(30,-20)
  \Line(-20,20)(-20,-20)
  \Line(-20,20)(20,20)
  \Line(-20,20)(20,0)
\Text(30,-3)[cb]{$P_7$}
  \SetWidth{1.6}
  \Line(20,20)(30,20)
  \Line(20,-20)(30,-20)
  \Line(20,20)(20,-20)
\end{picture}}
& \hspace*{16mm} = &\frac{1}{m^2} \sum_{i=-3}^{0} A_{i} \vep^i   + {\mathcal O}(\vep) \, ,
\end{eqnarray}

\begin{eqnarray} 
 A_{-3} &=& \frac{1}{32 (1+y)}\, , \nn \\
 A_{-2} &=&\frac{1}{16 (1+y)} \Bigl[
1 -  G(-1;y) \Bigr] \, , \nn \\
 A_{-1} &=&\frac{1}{16 (1+y) (x+1) } \Bigl[
  5 \zeta(2) x+2 x-3 \zeta(2)+2-2 (x+1) 
 G(-1;y)
 \nn \\
 & &
 +2 (x+1) G(-1,-1;y)+ (x-1) G(0,0;x)-2 (x-1) G(0,1;x) \Bigr] \, , \nn \\
 A_{0} &=& \frac{1}{16 (1+y) (x+1) (1-x+x^2 + x y) } \Bigl[8 (x+1) (x-1)^2 G(1;x) G(-1,-1;y) 
 \nn \\
 & &
 +4 (x+1)(x-1)^2 G(-1;y) G(1,0;x) 
-8 (x+1)(x-1)^2 G(-1;y) G(1,1;x) 
\nn \\
& &
-4 (x+1)(x-1)^2 G(1,1,0;x) 
+8 (x+1) (x-1)^2 G(1,1,1;x) 
\nn \\
& &
- 8 \left(x^2 \! +\! y x\! -\! x\! +\! 1\right) (x-1) \zeta(2) G(-1;x) 
+ 3 \left(3 x^2\! +\! y x-\! x\! -\! 1\right)(x-1) \zeta(2) G(0;x) 
\nn \\
& &
- 4 \left(x^2\! +\! 4 y x\! -\! 4 x\! +\! 7\right) (x-1)\zeta(2) G(1;x) 
+ 8 (y x\! -\! x\! +\! 2) (x-1)\zeta(2) G\left(-1/y;x\right)
\nn \\
& &
+ 4 (y x-x+2) (x-1) G\left(-1/y;x\right) G(-1,-1;y) -4 x (2 x+y-1) \times
\nn \\
& &
 \times(x-1) G(-y;x) 
G(-1,-1;y) -2 x (2 x+y-1)(x-1) G(0;x) G(0,-1;y) 
\nn \\
& &
-2 (y x-x+2) (x-1)
G\left(-1/y;x\right) G(0,-1;y) +2 x (2 x+y-1)  (x-1) \times
\nn \\
& &
\times G(-y;x) G(0,-1;y)+2 \left(x^2 \! +\! y x\! -\! x\! +\! 1\right) (x-1)
 G(0,0;x) -2 
\left(x^2\! +\! y x\! -\! x\! +\! 1\right)\times
\nn \\
& &
\times (x-1)G(-1;y) G(0,0;x) -4 \left(x^2+y x-x+1\right)(x-1) G(0,1;x) 
\nn \\
& &
+4 \left(x^2+y x-x+1\right) (x-1)G(-1;y) G(0,1;x) 
+2 (y x-x+2)(x-1) G(-1;y)
\nn \\
& &
 G\left(-1/y,0;x\right) -4 (y 
x-x+2) (x-1) G(-1;y) G\left(-1/y,1;x\right)
\nn \\
& &
-2 x (2 x+y-1) (x-1)
G(-1;y) G(-y,0;x) +4 x (2 x+y-1) (x-1) G(-1;y) \times
\nn \\
& &
\times G(-y,1;x)-2 
\left(x^2+y x-x+1\right) (x-1) G(-1,0,0;x)+4 \left(x^2+y x-x+1\right)
\times
\nn \\
& &
\times (x-1) 
G(-1,0,1;x)+2 x (2 x+y-1)(x-1) G(0,0,-1;y) + (5 x^2+3 y 
x
\nn \\
& &
-3 x+1)  (x-1)G(0,0,0;x)-2 \left(5 x^2+3 y x-3 x+1\right)  (x-1)
G(0,0,1;x)
\nn \\
& &-2 \left(x^2+y x-x+1\right) 
(x-1) G(0,1,0;x)+4 
\left(x^2+y x-x+1\right) (x-1) G(0,1,1;x)
\nn \\
& &
-2 \left(x^2+2 y x-2 x+3\right) (x-1) G(1,0,0;x)+4 (x^2+2 y x-2 x
\nn \\
& &
+3)(x-1) G(1,0,1;x) 
+2 (y x-x+2) G\left(-1/y,0,0;x\right) (x-1)-4 (y x-x
\nn \\
& &
+2) (x-1)
G\left(-1/y,0,1;x\right) -2 (y x-x+2) (x-1)
G\left(-1/y,1,0;x\right) +4 (y x-x
\nn \\
& &
+2) (x-1)
G\left(-1/y,1,1;x\right) +2 x (2 x+y
-1) (x-1)G(-y,1,0;x) 
-4 x (2 x+y
\nn \\
& &-1)(x-1) G(-y,1,1;x) +2 \bigl(5 \zeta(2) x^3+4 
\zeta(3) x^3+2 x^3+2 y x^2+5 y \zeta(2) x^2
\nn \\
& &
-8 \zeta(2) x^2 \! + \! 5 y 
\zeta(3) x^2 \! -4 \zeta(3) x^2 \! +2 y x-3 y \zeta(2) x+8 \zeta(2) x+6 
\zeta(3) x-3 \zeta(2)-\zeta(3)
\nn \\
& &
+2\bigr)-2 \bigl(5 \zeta(2) x^3+2 
x^3+2 y x^2+4 y \zeta(2) x^2-9 \zeta(2) x^2+2 y x-4 y \zeta(2) x+7 
\zeta(2) x
\nn \\
& &
-3 \zeta(2)+2\bigr) G(-1;y)+4 (x+1) \left(x^2+y x-x+1\right) G(-1,-1;y)-4 (x+1) \bigl(x^2
\nn \\
& &
+y x-x+1\bigr) G(-1,-1,-1;y)+2 x 
(x+1) (y+1) G(-1,0,-1;y)-2 \bigl(5 x^3+3 y x^2
\nn \\
& &
-6 x^2-y x+2 x+1\bigr) 
G(0,-1,-1;y) \Bigl] \, . 
 \end{eqnarray}

The topology in Fig.~\ref{nlMIs}-(e) involves two MIs. One of them is 

\begin{eqnarray}
\hbox{\begin{picture}(0,0)(0,0)
\SetScale{1}
  \SetWidth{.5}
  \Line(-30,20)(30,20)
  \Line(-30,-20)(30,-20)
  \Line(-20,20)(-20,-20)
  \Line(-20,20)(20,20)
  \Line(-20,0)(20,0)
  \SetWidth{1.6}
  \Line(20,20)(30,20)
  \Line(20,-20)(30,-20)
  \Line(20,20)(20,-20)
\end{picture}}
& \hspace*{9mm} \!\!=\!\!\! & \int\! \!\! \frac{{\mathfrak
D}^dk_1  {\mathfrak D}^dk_2}{\!P_0\!\left(\!k_2 \!\right) 
\!P_0\!\left(\!k_1\!-\!\!k_2\!\right)
\!P_0\!\left(\!k_1\!-\!\!p_1\!\right) 
\!P_0\!\left(\!k_2\!-\!\!p_1\!\right) 
\!P_0\!\left(\!k_1\!-\!\!p_1\!-\!\!p_2\!\right)
\!P_m\!\left(\!k_1\!-\!\!p_3\!\right) 
\!P_m\!\left(\!k_2\!-\!\!p_3\!\right)} ,
\end{eqnarray} 

\vspace*{5mm}
\noindent where

\begin{eqnarray}
\hbox{\begin{picture}(0,0)(0,0)
\SetScale{1}
  \SetWidth{.5}
  \Line(-30,20)(30,20)
  \Line(-30,-20)(30,-20)
  \Line(-20,20)(-20,-20)
  \Line(-20,20)(20,20)
  \Line(-20,0)(20,0)
  \SetWidth{1.6}
  \Line(20,20)(30,20)
  \Line(20,-20)(30,-20)
  \Line(20,20)(20,-20)
\end{picture}}
& \hspace*{16mm} = &\frac{1}{m^6} \sum_{i=-4}^{-1} A_{i} \vep^i   + {\mathcal O}(\vep^0) \, ,
\end{eqnarray}

\begin{eqnarray} 
A_{-4} &=& \frac{x}{12 (1-x)^2 (1+y)^2} \, , \nn \\
A_{-3} &=&\frac{x}{96 (1-x)^2 (1+y)^2} \Bigl[ -14 G(-1;y)+9 G(0;x)-18 G(1;x)\Bigr] \, , \nn \\
A_{-2} &=&\frac{x}{48 (1-x)^2 (1+y)^2} \Bigl[ -22  \zeta(2)-12  G(-1;y) G(0;x)+24  G(-1;y) G(1;x) \nn \\
& & +4 G(-1,-1;y)+3  G(0,0;x)-6  G(0,1;x)-6  G(1,0;x)+12  G(1,1;x) \Bigr] \, , \nn \\
 A_{-1} &=&\frac{x}{48 (1-x)^2 (1+y)^2} \Bigl[  -29  \zeta(3)+46  \zeta(2) G(-1;y)-21  \zeta(2) G(0;x)
\nn \\
& & 
 +66 
\zeta(2) G(1;x)-24  \zeta(2) G\left(-1/y;x\right)+24  
G(0;x) G(-1,-1;y)
\nn \\
& &
-24 G(1;x) G(-1,-1;y)-12  G\left(-1/y;x\right) G(-1,-1;y)
\nn \\
& &
-12  G(-y;x) G(-1,-1;y)-6  G(0;x) G(0,-1;y)+6  G
\left(-1/y;x\right) G(0,-1;y)
\nn \\
& &
+6  G(-y;x) G(0,-1;y)-12  
G(-1;y) G(0,0;x)+24  G(-1;y) G(0,1;x)
\nn \\
& &
+36  G(-1;y) G(1,0;x)-72  
G(-1;y) G(1,1;x)-6  G(-1;y) G\left(-1/y,0;x\right)
\nn \\
& &
+12  
G(-1;y) G\left(-1/y,1;x\right)-6  G(-1;y) G(-y,0;x)+12  
G(-1;y) G(-y,1;x)
\nn \\
& &
+32  G(-1,-1,-1;y)-12  G(-1,0,-1;y)-12  
G(0,-1,-1;y)
\nn \\
& &
+6  G(0,0,-1;y)+6  G(1,0,0;x)-12  G(1,0,1;x)-12  
G(1,1,0;x)
\nn \\
& &
+24  G(1,1,1;x)-6  G\left(-1/y,0,0;x\right)+12  
G\left(-1/y,0,1;x\right)+6  G\left(-1/y,1,0;x\right)
\nn \\
& &
-12  G\left(-1/y,1,1;x\right)+6  G(-y,1,0;x)-12  
G(-y,1,1;x)\Bigr] \, .
\end{eqnarray} 

The second MI for the topology in  Fig.~\ref{nlMIs}-(e) is

\begin{eqnarray}
\hbox{\begin{picture}(0,0)(0,0)
\SetScale{1}
  \SetWidth{.5}
  \Line(-30,20)(30,20)
  \Line(-30,-20)(30,-20)
  \Line(-20,20)(-20,-20)
  \Line(-20,20)(20,20)
  \Line(-20,0)(20,0)
\Text(30,-3)[cb]{$P_1$}
  \SetWidth{1.6}
  \Line(20,20)(30,20)
  \Line(20,-20)(30,-20)
  \Line(20,20)(20,-20)
\end{picture}}
& \hspace*{12mm} \!\!=\!\!\! & \int\! \!\! \frac{{\mathfrak
D}^dk_1  {\mathfrak
D}^dk_2 \, k_1^2}{\!P_0\!\left(\!k_2 \!\right) 
\!P_0\!\left(\!k_1\!\!-\!\!k_2\!\right)
\!P_0\!\left(\!k_1\!\!-\!\!p_1\!\right) 
\!P_0\!\left(\!k_2\!\!-\!\!p_1\!\right) 
\!P_0\!\left(\!k_1\!\!-\!\!p_1\!\!-\!\!p_2\!\right)
\!P_m\!\left(\!k_1\!\!-\!\!p_3\!\right) 
\!P_m\!\left(\!k_2\!\!-\!\!p_3\!\right)} ,
\end{eqnarray} 

\vspace*{5mm}
\noindent where

\begin{eqnarray}
\hbox{\begin{picture}(0,0)(0,0)
\SetScale{1}
  \SetWidth{.5}
  \Line(-30,20)(30,20)
  \Line(-30,-20)(30,-20)
  \Line(-20,20)(-20,-20)
  \Line(-20,20)(20,20)
  \Line(-20,0)(20,0)
\Text(35,-3)[cb]{$P_1$}
  \SetWidth{1.6}
  \Line(20,20)(30,20)
  \Line(20,-20)(30,-20)
  \Line(20,20)(20,-20)
\end{picture}}
& \hspace*{16mm} = &\frac{1}{m^4} \sum_{i=-4}^{-1} A_{i} \vep^i   + {\mathcal O}(\vep^0) \, ,
\end{eqnarray}

\begin{eqnarray} 
A_{-4} &=& \frac{1}{24 (1+y)^2} \, , \nn \\
A_{-3} &=&\frac{1}{96 (1+y)^2} \Bigl[-10 G(-1;y)+3 G(0;x)-6 G(1;x) \Bigr]\, , \nn \\
A_{-2} &=&\frac{1}{192 (1+y)^2} \Bigl[ -47 \zeta(2)-24 G(-1;y) G(0;x)+48 G(-1;y) G(1;x)+32 G(-1,-1;y)
\nn \\
& & -6 G(0,-1;y) \Bigr]\, , \nn \\
A_{-1} &=&\frac{1}{192 (1+y)^2} \Bigl[ -85 \zeta(3)+188 \zeta(2) G(-1;y)+96 \zeta(2) G(1;x)-96 \zeta(2) \
G\left(-1/y;x\right)
\nn \\
& &
+96 G(0;x) G(-1,-1;y)-96 G(1;x) 
G(-1,-1;y)-48 G\left(-1/y;x\right) G(-1,-1;y)
\nn \\
& &
-48 G(-y;x) 
G(-1,-1;y)-24 G(0;x) G(0,-1;y)+24 G\left(-1/y;x\right) 
G(0,-1;y)
\nn \\
& &
+24 G(-y;x) G(0,-1;y)+48 G(-1;y) G(1,0;x)-96 G(-1;y) 
G(1,1;x)
\nn \\
& &
-24 G(-1;y) G\left(-1/y,0;x\right)+48 G(-1;y) 
G\left(-1/y,1;x\right)
-24 G(-1;y) G(-y,0;x)
\nn \\
& &+48 G(-1;y) 
G(-y,1;x)+64 G(-1,-1,-1;y)-24 G(-1,0,-1;y)
\nn \\
& &
-12 G(0,-1,-1;y)+6 
G(0,0,-1;y)+24 G(1,0,0;x)-48 G(1,0,1;x)
\nn \\
& &-48 G(1,1,0;x)+96 
G(1,1,1;x)-24 G\left(-1/y,0,0;x\right)+48 
G\left(-1/y,0,1;x\right)
\nn \\
& &
+ \! 24 G\left(-1/y,\! 1,0;x\right)\! -\! 48 G\left(-1/y,\! 1,\! 1;x\right)\! +\! 24 G(-y,\! 1,0;x)\! -\! 48 
G(-y,1,1;x) \Bigr] . 
\end{eqnarray}

Finally, the topology in  Fig.~\ref{nlMIs}-(f) involves three MIs. 
The first MI we chose is 

\begin{eqnarray}
\hbox{\begin{picture}(0,0)(0,0)
\SetScale{1}
  \SetWidth{.5}
  \Line(-30,20)(30,20)
  \Line(-30,-20)(30,-20)
  \Line(-20,20)(-20,-20)
  \Line(-20,20)(20,20)
  \Line(0,-20)(0,20)
  \SetWidth{1.6}
  \Line(20,20)(30,20)
  \Line(20,-20)(30,-20)
  \Line(20,20)(20,-20)
\end{picture}}
& \hspace*{9mm} \!\!=\!\!\! & \int\! \!\! \frac{{\mathfrak
D}^dk_1  {\mathfrak
D}^dk_2 \,}{\!P_0\!\left(\!k_1 \!\right) 
\!P_0\!\left(\!k_2\right)
\!P_0\!\left(\!k_1\!\!-\!k_2\!\right) 
\!P_0\!\left(\!k_2\!\!-\!p_1\!\right)  
\!P_0\!\left(\!k_1\!\!-\!p_1\!\!-\!\!p_2\!\right) 
\!P_0\!\left(\!k_2\!\!-\!p_1\!\!-\!\!p_2\!\right)
\!P_m\!\left(\!k_1\!\!-\!p_3\!\right)} ,
\end{eqnarray} 

\vspace*{5mm}
\noindent where

\begin{eqnarray}
\hbox{\begin{picture}(0,0)(0,0)
\SetScale{1}
  \SetWidth{.5}
  \Line(-30,20)(30,20)
  \Line(-30,-20)(30,-20)
  \Line(-20,20)(-20,-20)
  \Line(-20,20)(20,20)
  \Line(0,-20)(0,20)
  \SetWidth{1.6}
  \Line(20,20)(30,20)
  \Line(20,-20)(30,-20)
  \Line(20,20)(20,-20)
\end{picture}}
& \hspace*{16mm} = &\frac{1}{m^6} \sum_{i=-4}^{-1} A_{i} \vep^i   + {\mathcal O}(\vep^0) \, ,
\end{eqnarray}

\begin{eqnarray} 
 A_{-4} &=&  \frac{x^2}{24 (1-x)^4 (1+y)}  \, , \nn \\
 A_{-3} &=&  \frac{x^2}{96 (1-x)^4 (1+y)} \Bigl[-10 G(-1;y)+3 G(0;x)-6 G(1;x)  \Bigr] \, , \nn \\
 A_{-2} &=&  \frac{x^2}{48 (1-x)^4 (1+y)}\! \Bigl[ -5 \zeta(2)-6 G(\!-1;y) G(0;x)\!+\!12 G(\!-1;y) G(1;x)\!+\!8 G(\!-1,\!-1;y)\!  \Bigr] \, , \nn \\
 A_{-1} &=&  \frac{x^2}{48 (1-x)^4 (1+y)} \Bigl[-13 \zeta(3) +38 \zeta(2) G(-1;y)+9 \zeta(2) G(0;x)+6 \zeta(2)
G(1;x)
\nn \\
& &
-24 \zeta(2) G\left(-1/y;x\right)+24 G(0;x) 
G(-1,-1;y)-24 G(1;x) G(-1,-1;y)
\nn \\
& &
-12 G\left(-1/y;x\right) 
G(-1,-1;y)-12 G(-y;x) G(-1,-1;y)-6 G(0;x) G(0,-1;y)
\nn \\
& &
+6 
G\left(-1/y;x\right) G(0,-1;y)+6 G(-y;x) G(0,-1;y)+12 
G(-1;y) G(1,0;x)
\nn \\
& &
-24 G(-1;y) G(1,1;x)-6 G(-1;y) 
G\left(-1/y,0;x\right)+12 G(-1;y) G\left(-1/y,1;x\right)
\nn \\
& &
-6 G(-1;y) G(-y,0;x)+12 G(-1;y) G(-y,1;x)+16 G(-1,-1,-1;y)
\nn \\
& &-12 
G(-1,0,-1;y)-12 G(0,-1,-1;y)+6 G(0,0,-1;y)+6 G(1,0,0;x)
\nn \\
& &-12 
G(1,0,1;x)-12 G(1,1,0;x)+24 G(1,1,1;x)-6 G\left(-1/y,0,0;x\right)
\nn \\
& &
+12 G\left(-1/y,0,1;x\right)+6
G\left(-1/y,1,0;x\right)-12 G\left(-1/y,1,1;x\right)+6 G(-y,1,0;x)
\nn \\
& &-12 G(-y,1,1;x)  \Bigr] \, .
\end{eqnarray} 

The second MI for the topology Fig.~\ref{nlMIs}-(f) was chosen as follows 

\begin{eqnarray}
\hbox{\begin{picture}(0,0)(0,0)
\SetScale{1}
  \SetWidth{.5}
  \Line(-30,20)(30,20)
  \Line(-30,-20)(30,-20)
  \Line(-20,20)(-20,-20)
  \Line(-20,20)(20,20)
  \Line(0,-20)(0,20)
\Text(30,-3)[cb]{$P_4$}
  \SetWidth{1.6}
  \Line(20,20)(30,20)
  \Line(20,-20)(30,-20)
  \Line(20,20)(20,-20)
\end{picture}}
& \hspace*{12mm} \!\!=\!\!\! & \int\! \!\! \frac{{\mathfrak
D}^dk_1  {\mathfrak
D}^dk_2 \, \left( k_1 -p_1\right)^2}{\!P_0\!\left(\!k_1 \!\right) 
\!P_0\!\left(\!k_2\right)
\!P_0\!\left(\!k_1\!\!-\!\!k_2\!\right) 
\!P_0\!\left(\!k_2\!\!-\!\!p_1\!\right)  
\!P_0\!\left(\!k_1\!\!-\!\!p_1\!\!-\!\!p_2\!\right) 
\!P_0\!\left(\!k_2\!\!-\!\!p_1\!\!-\!\!p_2\!\right)
\!P_m\!\left(\!k_1\!\!-\!\!p_3\!\right)} ,
\end{eqnarray} 

\vspace*{5mm}
\noindent where

\begin{eqnarray}
\hbox{\begin{picture}(0,0)(0,0)
\SetScale{1}
  \SetWidth{.5}
  \Line(-30,20)(30,20)
  \Line(-30,-20)(30,-20)
  \Line(-20,20)(-20,-20)
  \Line(-20,20)(20,20)
  \Line(0,-20)(0,20)
\Text(30,-3)[cb]{$P_4$}
  \SetWidth{1.6}
  \Line(20,20)(30,20)
  \Line(20,-20)(30,-20)
  \Line(20,20)(20,-20)
\end{picture}}
& \hspace*{16mm} = &\frac{1}{m^4} \sum_{i=-2}^{-1} A_{i} \vep^i   + {\mathcal O}(\vep^0) \, ,
\end{eqnarray}

\begin{eqnarray} 
 A_{-2} &=& \frac{x^2}{16 (1-x)^3 (1+x)} \Bigl[4 \zeta(2) +G(0,0;x) -2 G(0,1;x) \Bigr] \, , \nn \\
 A_{-1} &=& \frac{x^2}{16 (1-x)^3 (1+x)} \Bigl[ 5 \zeta(3)+24 \zeta(2) G(-1;x) -\zeta(2) G(0;x) -8 \
\zeta(2) G(1;x) 
\nn \\
& &
-8 \zeta(2) G\left(-1/y;x\right) -4 G
\left(-1/y;x\right) G(-1,-1;y) +4 G(-y;x) G(-1,-1;y) 
\nn \\
& &+2 
G(0;x) G(0,-1;y) +2 G\left(-1/y;x\right) G(0,-1;y) -2 
G(-y;x) G(0,-1;y) 
\nn \\
& &
-2 G(-1;y) G\left(-1/y,0;x\right) +4 
G(-1;y) G\left(-1/y,1;x\right) +2 G(-1;y) G(-y,0;x)
\nn \\
& &
 -4 
G(-1;y) G(-y,1;x) +6 G(-1,0,0;x) -12 G(-1,0,1;x) +4 
G(0,-1,-1;y) 
\nn \\
& &
-2 G(0,0,-1;y) +2 G(0,0,0;x) -4 G(0,0,1;x) 
-6 G(0,1,0;x) +12 G(0,1,1;x)
\nn \\
& & -2 G(1,0,0;x) +4 G(1,0,1;x) 
-2 G\left(-1/y,0,0;x\right) +4 
G\left(-1/y,0,1;x\right) 
\nn \\
& &
+ 2 G\left(-1/y,1,0;x\right) \! - \! 4 G\left(-1/y,1,1;x\right) \! 
-\! 2 G(-y,1,0;x) \! +\! 4 G(-y,1,1;x) \Bigr] . 
\end{eqnarray}

The last MI for topology~\ref{nlMIs}-(f) is 

\begin{eqnarray}
\hbox{\begin{picture}(0,0)(0,0)
\SetScale{1}
  \SetWidth{.5}
  \Line(-30,20)(30,20)
  \Line(-30,-20)(30,-20)
  \Line(-20,20)(-20,-20)
  \Line(-20,20)(20,20)
  \Line(0,-20)(0,20)
\Text(30,-3)[cb]{$P_9$}
  \SetWidth{1.6}
  \Line(20,20)(30,20)
  \Line(20,-20)(30,-20)
  \Line(20,20)(20,-20)
\end{picture}}
& \hspace*{12mm} \!\!=\!\!\! & \int\! \!\! \frac{{\mathfrak
D}^dk_1  {\mathfrak
D}^dk_2 \, \left[\left( k_2 -p_3\right)^2 + m^2 \right]}{
\!P_0\!\left(\!k_1 \!\right) 
\!P_0\!\left(\!k_2\!\right)
\!P_0\!\left(\!k_1\!\!-\!\!k_2\!\right) 
\!P_0\!\left(\!k_2\!\!-\!\!p_1\!\right)  
\!P_0\!\left(\!k_1\!\!-\!\!p_1\!\!-\!\!p_2\!\right) 
\!P_0\!\left(\!k_2\!\!-\!\!p_1\!\!-\!\!p_2\!\right)
\!P_m\!\left(\!k_1\!\!-\!\!p_3\!\right)} ,
\end{eqnarray} 

\vspace*{5mm}
\noindent with

\begin{eqnarray}
\hbox{\begin{picture}(0,0)(0,0)
\SetScale{1}
  \SetWidth{.5}
  \Line(-30,20)(30,20)
  \Line(-30,-20)(30,-20)
  \Line(-20,20)(-20,-20)
  \Line(-20,20)(20,20)
  \Line(0,-20)(0,20)
\Text(30,-3)[cb]{$P_9$}
  \SetWidth{1.6}
  \Line(20,20)(30,20)
  \Line(20,-20)(30,-20)
  \Line(20,20)(20,-20)
\end{picture}}
& \hspace*{16mm} = &\frac{1}{m^4} \sum_{i=-4}^{-1} A_{i} \vep^i   + {\mathcal O}(\vep^0) \, ,
\end{eqnarray}

\begin{eqnarray} 
 A_{-4} &=& \frac{7 x^2}{192 (1-x)^4} \, , \nn \\
 A_{-3} &=& \frac{x^2}{96 (1-x)^4} \Bigl[ -8 G(-1;y)+3 G(0;x)-6 G(1;x)\Bigr]\, , \nn \\
 A_{-2} &=& \frac{x^2}{24 (1-x)^4} \Bigl[ -5 \zeta(2)-3 G(-1;y) G(0;x)+6 G(-1;y) G(1;x)+2 G(-1,-1;y) \Bigr]\, , \nn \\
 A_{-1} &=& \frac{x^2}{48 (1-x)^4} \Bigl[-29 \zeta(3)+58 \zeta(2) G(-1;y)+9 \zeta(2) G(0;x)+6 \zeta(2) \
G(1;x)
\nn \\
& &
-24 \zeta(2) G\left(-1/y;x\right)+24 G(0;x) 
G(-1,-1;y)-24 G(1;x) G(-1,-1;y)
\nn \\
& &
-12 G\left(-1/y;x\right) 
G(-1,-1;y)-12 G(-y;x) G(-1,-1;y)-6 G(0;x) G(0,-1;y)
\nn \\
& &+6 
G\left(-1/y;x\right) G(0,-1;y)+6 G(-y;x) G(0,-1;y)+12 
G(-1;y) G(1,0;x)
\nn \\
& &
-24 G(-1;y) G(1,1;x)-6 G(-1;y) 
G\left(-1/y,0;x\right)+12 G(-1;y) G\left(-1/y,1;x\right)
\nn \\
& &
-6 G(-1;y) G(-y,0;x)+12 G(-1;y) G(-y,1;x)+32 G(-1,-1,-1;y)
\nn \\
& &
-12 
G(-1,0,-1;y)-12 G(0,-1,-1;y)+6 G(0,0,-1;y)+6 G(1,0,0;x)
\nn \\
& &
-12 
G(1,0,1;x)-12 G(1,1,0;x)+24 G(1,1,1;x)-6 G\left(-1/y,0,0;x\right)
\nn \\
& &
+12 G\left(-1/y,0,1;x\right)+6 
G\left(-1/y,1,0;x\right)-12 G\left(-1/y,1,1;x\right)+6 G(-y,1,0;x)
\nn \\
& &
-12 G(-y,1,1;x) \Bigr]\, . 
\end{eqnarray}


\end{document}